\newif\ifAddLineNumbers
\newif\ifAddVersion
\def\draftVersion{6.2}
\newif\ifDraftWatermark
\DraftWatermarkfalse

\newif\ifUseTwoColElsArt
\UseTwoColElsArttrue

\ifUseTwoColElsArt
\documentclass[twocolumn, 3p]{elsarticle}
\else
\documentclass[preprint]{elsarticle}
\fi

\usepackage{dcolumn}% Align table columns on decimal point
\usepackage[linkbordercolor={0 .5 0},
            citebordercolor={.8 0 0},
            urlbordercolor={0 0 .5},
            raiselinks=true,
            breaklinks=true,
            bookmarks=false,
            pdfborder={0 0 0.5 [2]}]{hyperref}

\makeatletter
\providecommand{\doi}[1]{%
  \begingroup
    \let\bibinfo\@secondoftwo
    \urlstyle{rm}%
    \href{http://dx.doi.org/#1}{%
      doi:\discretionary{}{}{}%
      \nolinkurl{#1}%
    }%
  \endgroup
}
\makeatother

\usepackage{color}
\definecolor{carmine}{rgb}{0.59, 0.0, 0.09}
\definecolor{mediumpersianblue}{rgb}{0.0, 0.4, 0.65}
\definecolor{green(html/cssgreen)}{rgb}{0.0, 0.5, 0.0}
\usepackage{listings}
\ifAddLineNumbers
\usepackage{lineno}
\linenumbers
\fi

\ifDraftWatermark
\usepackage{type1cm}
\usepackage{eso-pic}
\makeatletter
\AddToShipoutPicture{%
  \setlength{\@tempdimb}{.5\paperwidth}%
  \setlength{\@tempdimc}{.5\paperheight}%
  \setlength{\unitlength}{1pt}%
  \put(\strip@pt\@tempdimb,\strip@pt\@tempdimc){%
    \makebox(6,-7){\rotatebox{45}{\textcolor[gray]{0.885}%
        {\fontsize{6cm}{6cm}\selectfont{DRAFT}}}}%
  }%
}
\makeatother
\fi

\def\arianna      {ARIANNA}
\def\cheren       {Cherenkov}
\def\myname       {C.Reed}

\def\fluxUpLim    {${1.9\!\!\times\!\!10^{-23}}$~$\gev^{-1}\:\cm^{-2}\:\second^{-1}\:\sr^{-1}$}
\def\fluxEnBin    {$10^{8.5}-10^{9.5}$~{\gev}}

\def\sctitle      {A First Search for Cosmogenic Neutrinos with the ARIANNA
Hexagonal Radio Array}

\def\etal         {\textit{et~al.}}

\newcommand {\veff}     {\ensuremath{V_{\mathit{eff}}}}
\newcommand {\lint}     {\ensuremath{L_{\mathit{int}}}}
\newcommand {\tlive}    {\ensuremath{t_{\mathit{live}}}}
\newcommand {\emin}     {\ensuremath{E_{\mathit{min}}}}
\newcommand {\emax}     {\ensuremath{E_{\mathit{max}}}}
\newcommand {\erec}     {\ensuremath{E_{\mathit{rec}}}}
\newcommand {\esh}      {\ensuremath{E_{\mathit{sh}}}}

\newcommand {\fig}[1]{Fig.~\ref{#1}}
\newcommand {\figs}[2]{Figs.~\ref{#1} and~\ref{#2}}
\newcommand {\Fig}[1]{Figure~\ref{#1}}
\newcommand {\eq}[1]{Eq.~\ref{#1}}

\newcommand {\sect}[1]{Sect.~\ref{#1}}
\newcommand {\Sect}[1]{Section~\ref{#1}}

\newcommand {\tab}[1]{Table~\ref{#1}}

\newcommand {\ord}      {\ensuremath{\mathcal{O}}}

\newcommand {\dg}    {\mbox{$^\circ$}}
\newcommand {\db}    {\mbox{${\rm dB}$}}
\newcommand {\sr}    {\mbox{${\rm sr}$}}

\newcommand {\pev}   {\mbox{${\rm PeV}$}}
\newcommand {\tev}   {\mbox{${\rm TeV}$}}
\newcommand {\gev}   {\mbox{${\rm GeV}$}}

\newcommand {\mom}   {\mbox{\rm GeV$\kern-0.15em /\kern-0.12em c$}}
\newcommand {\mmom}  {\mbox{\rm MeV$\kern-0.15em /\kern-0.12em c$}}
\newcommand {\mass}  {\mbox{\rm GeV$\kern-0.15em /\kern-0.12em c^2$}}
\newcommand {\mmass} {\mbox{\rm MeV$\kern-0.15em /\kern-0.12em c^2$}}

\newcommand {\km}    {\mbox{${\rm km}$}}
\newcommand {\m}     {\mbox{${\rm m}$}}

\newcommand {\cm}    {\mbox{${\rm cm}$}}

\newcommand {\watt}  {\mbox{${\rm W}$}}

\newcommand {\ns}    {\mbox{${\rm ns}$}}
\newcommand {\ps}    {\mbox{${\rm ps}$}}
\newcommand {\second}{\mbox{${\rm s}$}}

\newcommand {\hz}    {\mbox{${\rm Hz}$}}
\newcommand {\mhz}   {\mbox{${\rm MHz}$}}
\newcommand {\ghz}   {\mbox{${\rm GHz}$}}

\newcommand {\mpc}   {\mbox{${\rm Mpc}$}}
\newcommand {\knots} {\mbox{${\rm knots}$}}

\pdfoutput=1        % we are running pdflatex
\pdfcatalog{/PageMode(/UseOutlines)} % or UseNone

\begin{document}

\title{\sctitle}
% Authors for the 2014 HRA Science paper
% Includes hardware from 2013 and 2014 seasons

\author[uciPhys]{S.W.Barwick}
\author[uciPhys]{E.C.Berg}
\author[kansas,moscow]{D.Z.Besson}
\author[lbl,ucb]{G.Binder}
\author[washu]{W.R.Binns}
\author[uppsala]{D.Boersma}
\author[washu]{R.G.Bose}
\author[washu]{D.L.Braun}
\author[washu]{J.H.Buckley}
\author[washu]{V.Bugaev}
\author[nijmegen]{S.Buitink}
%\author[aarhus]{P.B.Christensen}
\author[uciPhys]{K.Dookayka}
\author[washu]{P.F.Dowkontt}
\author[uciPhys]{T.Duffin}
\author[uppsala]{S.Euler}
\author[lbl]{L.Gerhardt}
\author[uppsala]{L.Gustafsson}
\author[uppsala]{A.Hallgren}
\author[kansas,uciPhys]{J.C.Hanson}
\author[washu]{M.H.Israel}
\author[stony]{J.Kiryluk}
\author[lbl,ucb]{S.Klein}
\author[uciEE]{S.Kleinfelder}
%\author[aarhus]{R.E.Mikkelsen}
\author[stony]{H.Niederhausen}
\author[washu]{M.A.Olevitch}
\author[uciPhys]{C.Persichelli}
%\author[aarhus]{T.Poulsen}
\author[kansasIDL]{K.Ratzlaff}
\author[washu]{B.F.Rauch}
\author[uciPhys]{C.Reed\corref{cor1}}
\ead{cjreed@uci.edu}
\author[uciEE]{M.Roumi}
\author[uciEE]{A.Samanta}
\author[washu]{G.E.Simburger}
\author[lbl]{T.Stezelberger}
\author[uciPhys,uw]{J.Tatar\corref{cor1}}
\ead{jtatar@uci.edu}
\author[aarhus]{U.Uggerhoj}
\author[uciPhys]{J.Walker}
\author[uciPhys]{G.Yodh}
\author[kansasIDL]{R.Young}

\author{{\newline}{\newline}(The ARIANNA Collaboration)}

\address[uciPhys]{Dept. of Physics and Astronomy, University of California, Irvine, USA}
\address[uciEE]{Dept. of Electrical Engineering and Computer Science, University of California, Irvine, USA}
\address[kansas]{Dept. of Physics and Astronomy, University of Kansas, USA}
\address[kansasIDL]{Instrumentation Design Lab, University of Kansas, USA}
\address[lbl]{Lawrence Berkeley National Laboratory, USA}
\address[ucb]{Dept. of Physics, University of California, Berkeley, USA}
\address[moscow]{Moscow Engineering and Physics Institute, Russia}
\address[uw]{Center for Experimental Nuclear Physics and Astrophysics, University of Washington, USA}
\address[stony]{Stonybrook University, USA}
\address[uppsala]{University of Uppsala, Sweden}
\address[nijmegen]{University of Nijmegen, The Netherlands}
\address[aarhus]{Aarhus University, Denmark}
\address[washu]{Dept. of Physics and McDonnell Center for the Space Sciences, Washington University, St. Louis, USA}

\cortext[cor1]{Corresponding author}
\begin{abstract}\noindent
The {\arianna} experiment seeks to observe the diffuse flux of
neutrinos in the $10^{8}-10^{10}$~{\gev} energy range using a grid of
radio detectors at the surface of the Ross Ice Shelf of
Antarctica. The detector measures the coherent {\cheren} radiation
produced at radio frequencies, from about 100~{\mhz} to 1~{\ghz}, by
charged particle showers generated by neutrino interactions in the
ice. The {\arianna} Hexagonal Radio Array (HRA) is being constructed
as a prototype for the full array. During the 2013-14 austral summer,
three HRA stations collected radio data which was wirelessly
transmitted off site in nearly real-time. The performance of these
stations is described and a simple analysis to search for neutrino
signals is presented. The analysis employs a set of three cuts that
reject background triggers while preserving 90\% of simulated
cosmogenic neutrino triggers. No neutrino candidates are found in the
data and a model-independent 90\% confidence level Neyman upper limit
is placed on the all flavor ${\nu+\bar{\nu}}$ flux in a sliding
decade-wide energy bin.  The limit reaches a minimum of {\fluxUpLim}
in the {\fluxEnBin} energy bin. Simulations of the performance of the
full detector are also described. The sensitivity of the full
{\arianna} experiment is presented and compared with current neutrino
flux models.
\end{abstract}
\ifAddVersion
\else
\fi
\maketitle

%\comm{add Stecker's idea to test Lorentz invariance}

\section{Introduction}
\label{intro}

While the flux of cosmic rays has been measured to energies greater
than $10^{10}$~{\gev}~\cite{LetessierSelvon:2011dy}, the sources of
such high energy particles remain a mystery. No known galactic source
could accelerate particles to such energies, and no particular sources
of the very highest energy particles, with large rigidities, have been
found~\cite{Abreu:2012ybu, Weiler:2000ku, Sigl:2002yk,
  Sigl:2000sp}. Potential sources of such ultra-high energy (UHE)
cosmic rays are limited to our local supercluster (within about
50~{\mpc}) due to their interaction with the cosmic microwave
background (CMB)~\cite{Greisen:1966jv, Zatsepin:1966jv}. The mesons
produced by this process promptly decay to leptons, leading to a flux
of UHE neutrinos~\cite{Stecker:1973sy, Beresinsky:1969qj,
  Berezinsky:1975zz}.

%% A measurement of this UHE ``cosmogenic'' neutrino spectrum would
%% contribute significantly to the understanding of cosmic ray sources.
%% The shape of the spectrum can help determine how cosmic ray production
%% depends on source evolution and the injection
%% spectrum~\cite{Seckel:2005cm}. The overall normalization of the
%% neutrino flux can help constrain both the source evolution and the
%% cosmic ray composition. 

Cosmogenic neutrinos may reveal cosmic accelerators beyond our local
supercluster, as the mean free path of neutrinos through the CMB is
larger than the visible universe. Such neutrinos would be produced
within about 50~{\mpc} of the cosmic ray sources and would travel to
Earth without deflection by magnetic fields, potentially pointing back
to the accelerating objects.

Several large projects (AMANDA~\cite{Ahrens:2003pv},
IceCube~\cite{Aartsen:2013dsm, Halzen:2010yj},
ANITA~\cite{Gorham:2008yk, PhysRevD.82.022004, Gorham:2010kv} and
RICE~\cite{Kravchenko:2002mm,Kravchenko:2006qc}) exploit the fact that
ice is transparent to {\cheren} radiation (at both optical and radio
wavelengths) in order to search for cosmogenic neutrinos. These
experiments complement cosmogenic neutrino searches by air shower
detectors such as the Pierre Auger Observatory~\cite{Abraham:2007rj,
  Abraham:2009uy} and HiRes~\cite{Abbasi:2008hr, Martens:2007ff}.
Below energies of $10^{10}$~{\gev}, IceCube currently provides the
best constraints on the UHE neutrino flux and in the
$10^{4}-10^{6}$~{\gev} range, IceCube has observed an
extra-terrestrial diffuse neutrino flux~\cite{Aartsen:2014gkd}.

A new generation of neutrino experiments is emerging with the efforts
of ARA~\cite{Allison:2011wk, Allison:2014kha},
GNO~\cite{GNOWhitePaper} and the Antarctic Ross Ice Shelf Antenna
Neutrino Array ({\arianna}, described in this paper). These
experiments seek to extend the neutrino flux measurements to
ultra-high energies by constructing radio {\cheren} detectors that are
orders of magnitude larger in effective sensitive volume than current
experiments using well-understood and inexpensive technology.
Preparation is underway for the next generation of ballon-borne
experiments as well~\cite{Gorham:2013qfa}, with efforts like that of
EVA~\cite{Gorham:2011mt}. A large number of models predict cosmogenic
neutrino fluxes that are measurable by such experiments with improved
sensitivity to neutrinos above $10^{8}$~{\gev}, particularly in the
$10^{8}-10^{10}$~{\gev} range.  See \sect{arianna:performance} for
examples of such models.

The {\arianna} and ARA experiments are proposing the construction of
arrays of radio detectors in Antarctica that will reach effective
volumes {\ord}(100)~{$\km^3$}. A third radio experiment, GNO, is
exploring the construction of a radio neutrino telescope in
Greenland. These experiments will measure the radio-frequency (RF)
pulse emitted by the charged particle shower resulting from a UHE
neutrino interaction in ice via the Askaryan
effect~\cite{Askaryan:1962, Askaryan:1965}. The Askaryan radio pulse
has been measured in a variety of dielectric materials using particle
accelerators to induce charged particle
showers~\cite{Saltzberg:2000bk, Gorham:2006fy}.

%% \begin{figure}[t]
%%    \begin{center}
%%      \includegraphics[width=\linewidth]{img/Hra2013iceAttenVsReflect}
%%    \end{center}
%%    \caption{\label{intro:fig:attenVsR}
%%      The measured attenuation length and reflectivity parameter space.
%%      Fit contours are from data taken in 2011~\cite{AriannaIcePaper},
%%      while the broken lines represent measurements taken in
%%      2006~\cite{Barrella:2010vs}. The 530~{\mhz} line has been
%%      obtained from an interpolation of measurement data. Figure
%%      from Ref.~\cite{AriannaIcePaper}.}
%% \end{figure}

%% \begin{figure}[t]
%%    \begin{center}
%%      \includegraphics[width=\linewidth]{img/Hra2013iceAttenVsFreq}
%%    \end{center}
%%    \caption{\label{intro:fig:attenVsFreq}
%%      The attenuation length as a function of frequency from
%%      measurements made at the {\arianna} site in different years and
%%      with different radio antennas~\cite{AriannaIcePaper}.}
%% \end{figure}

The {\arianna} collaboration plans to construct a
$36\times36$~{$\km^2$} array of 1296 independent, autonomous radio
detector stations just below the surface of the Ross Ice Shelf. The
ice to water interface below the Ross Ice Shelf serves as a mirror for
radio waves, allowing the stations to observe neutrinos arriving from
the sky above the detector as well as from the horizon. The detector
will measure radio frequencies between about $100~{\mhz}-1~{\ghz}$.
This bandwidth is sensitive to the linear increase in power of the
Askaryan pulse with frequency up to $\sim\!1$~{\ghz} for signals
measured on the {\cheren} cone~\cite{AlvarezMuniz:2000fw}.

The {\arianna} site is roughly 100~{\km} from the McMurdo
Antarctic Station, which provides logistical support during
construction. Despite the relative closeness of McMurdo, the
{\arianna} site is free of anthropogenic RF noise due to its being
buffered by Minna Bluff to the north and the Transantarctic Mountains
to the west.

Properties of the ice at the {\arianna} site have been measured by
transmitting polarized radio pulses into the ice and observing the
reflected pulses at multiple locations. These measurements indicate
that the ice to water interface is a near perfect mirror. The
attenuation length, measured to be between 400 and 500~{\m} for radio
frequencies, is found to be comparable to the average thickness of the
Ross Ice Shelf. The ice shelf thickness has been measured to be
$576\pm8~{\m}$~\cite{AriannaIcePaper} including a firn layer within
the upper 60-70~{\m}~\cite{0034-4885-67-10-R03} (approximately). The
firn layer is characterized by a monotonic increase in mass density as
a function of depth. A more complete discussion of the ice properties
at the {\arianna} site is presented in Ref.~\cite{AriannaIcePaper}.

The construction of the {\arianna} Hexagonal Radio Array (HRA) is
approved for completion during the 2014-2015 austral summer. This
array of seven {\arianna} stations serves as a research and
development project for the full {\arianna}
array~\cite{AriannaIcrc2013}. Each HRA station consists of four
log-periodic dipole antennas (LPDAs), a high-speed data acquisition
(DAQ) system, wireless communication peripherals and local renewable
power generation.

The expected performance of the full {\arianna} telescope is presented
in \Sect{arianna}. The performance of the HRA stations is discussed
in \Sect{hra}. A first search for neutrino signals in the HRA data is
described in \Sect{ana}.

\section{The {\arianna} Telescope}
\label{arianna}

The {\arianna} experiment plans to measure the cosmogenic neutrino
flux using a large surface array of radio receivers. {\arianna} will
build upon previous UHE neutrino searches by greatly increasing the
size of the detector. This will improve the sensitivity to neutrinos
of $10^{8}-10^{10}$~{\gev} by a factor of 13 or more, depending on
model, relative to the current best limits (see
\sect{arianna:performance:nurates}). In order to maximize the
effective volume of the telescope, each {\arianna} station of the
$36\times36$ array will be separated from neighboring stations by
1~{\km}, so that a typical neutrino pulse will be observed by a single
station. The stations will measure the amplitude and direction of the
incoming radio pulse using multiple antennas, allowing the energy and
source direction of the primary neutrino to be determined.

The ice to water interface at the bottom of the Ross Ice Shelf
provides a near perfect mirror for radio
waves~\cite{AriannaIcePaper}. This allows the surface detectors to
measure reflected radio pulses produced by down-going neutrino-induced
showers, in addition to directly measuring the Askaryan radiation of
horizontal showers. As the Earth is opaque to UHE neutrinos, this
(greater than) $2\pi$~{\sr} solid angle acceptance contributes heavily
to the high sensitivity of the {\arianna} telescope. The sensitivity
of the experiment also benefits from a low energy threshold (below
$10^{8}$~{\gev}) and from the large number of detector stations made
possible by the ease of installation at the ice surface.

The flagship measurement of the {\arianna} telescope will be the
observation of the flux of cosmogenic neutrinos in the
$10^{8}-10^{10}$~{\gev} range.  The predicted flux of these neutrinos
depends on the chemical composition of UHE cosmic rays, the cosmic ray
injection spectrum and the cosmic ray source
evolution~\cite{Kampert:2012mx}. A measurement of the neutrino flux
will provide additional input to help constrain these parameters and
thus improve the understanding of both neutrino and cosmic ray
sources.

The observation of a significant number of neutrinos by {\arianna}
would allow further measurements to be performed. The shape of the
neutrino energy spectrum can help distinguish a flux due to strong
source evolution from one due to a soft injection
spectrum~\cite{Engel:2001hd, Seckel:2005cm}. A search for point-like
sources of UHE neutrinos will be a primary goal and has the potential
to reveal particle accelerators at distances beyond our local
supercluster. Further, the neutrino-nucleon cross section can be
measured at center of mass energies around 10~{\tev} through the
angular dependence of the
flux~\cite{Connolly:2011vc,Klein:2013xoa}. In addition, the flavor
composition of the neutrino flux may be
explored~\cite{Wang:2013njo}. Once the flux of neutrinos is known,
such observations may be improved by redeploying the surface detectors
in order to maximize angular and energy resolution.

Even the lack of a measurable neutrino flux would have profound
consequences. Such a scenario would imply that either the sources of
the highest energy cosmic rays are local, or that the sources have
astrophysically interesting properties, such as an iron-dominated
composition with a hard energy injection spectrum and an acceleration
energy cutoff below the photo-fragmentation
threshold~\cite{Olinto:2011ng}.

%% \begin{figure}[t]
%%    \begin{center}
%%      \includegraphics[width=\linewidth]{img/HraFourChanGeo}
%%    \end{center}
%%    \caption{\label{arianna:fig:fourChanGeo}
%%      The four radio antennas of an HRA station are oriented
%%      symmetrically around the DAQ box at a distance of 3~{\m}.
%%      %\comm{Get better graphic.}
%%    }
%% \end{figure}

\begin{figure}[t!]
   \begin{center}
     \includegraphics[width=\linewidth]{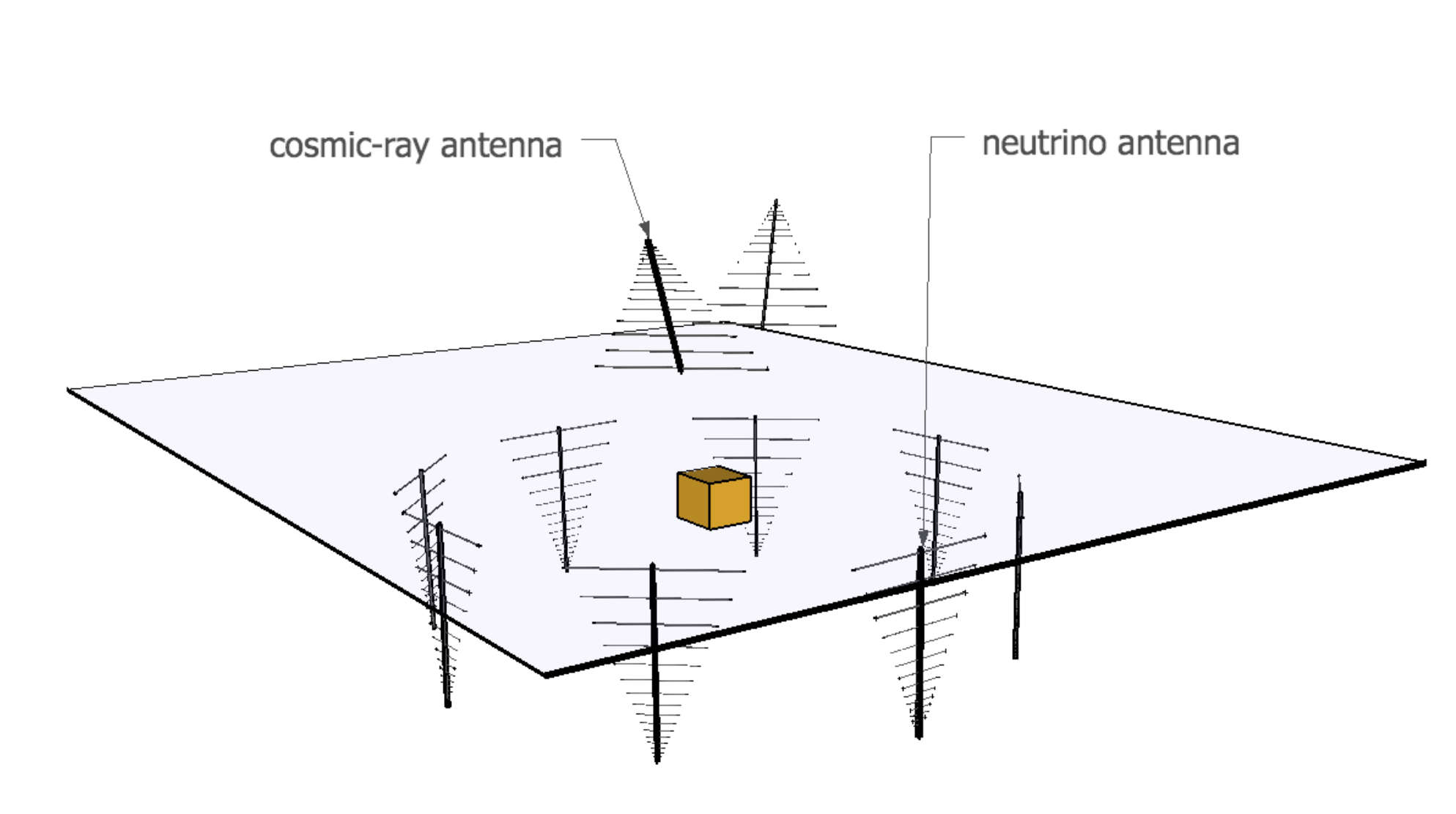}
   \end{center}
   \caption{\label{arianna:fig:CRChanGeo}
     An illustration of the antenna arrangement in a single {\arianna}
     station. For simulations of cosmic ray showers, all ten antennas
     are included in the simulation. For simulations of the detector
     response to neutrino signals, only the eight downward facing
     antennas are simulated. The stations built for the HRA detector
     use a smaller subset of only four downward facing antennas,
     arranged in a square.}
\end{figure}

The expected performance of the full {\arianna} experiment, described
in \sect{arianna:performance}, is determined through the simulation of
stations with eight downward facing LPDAs, as shown in
\fig{arianna:fig:CRChanGeo}. Studies of cosmic ray air showers suggest
that {\arianna} may trigger on radio pulses from such events. To that
end, the addition of two antennas that are directed upward at a
45{\dg} angle has been studied, as presented in
\sect{arianna:cosmics}.

\subsection{Simulation Methods}
\label{arianna:sims}

The performance of the {\arianna} telescope, detailed in
\sect{arianna:performance}, has been characterized by simulating the
production and detection of radio signals in the frequency domain.  An
additional simulation package, presented in \sect{arianna:sims:time},
has been developed to study the response of the detector in the time
domain. The analysis of HRA data, discussed in \sect{ana}, combines
both simulation tools in order to estimate expected neutrino
signals.

\subsubsection{Frequency Domain}
\label{arianna:sims:freq}

A set of simulation tools has been developed and used to calculate the
sensitivity and model the performance of the {\arianna} telescope. A
summary of the simulation is provided below, and further details may
be found in Ref.~\cite{KamleshPhd}. These tools simulate the
production of the Askaryan radio pulse resulting from neutrino
interactions and propagate it through the ice and firn to the
detector.

Neutrino interactions are simulated by forcing neutrinos to interact
within a fiducial volume and weighting the resulting events by the
probability with which they would occur. This probability accounts for
neutrinos lost to absorption within the atmosphere and the Earth's
crust, as well as tau neutrinos recovered due to $\nu_\tau$
regeneration effects~\cite{Halzen:1998be}. The neutrino-nucleon cross
section follows the parametrization presented in
Ref.~\cite{Gandhi:1998ri}, e.g.~$1.45\times10^{-32}$~{${\cm}^2$} at
$E_{\nu}=10^{9}$~{\gev}. As {\arianna} stations are designed to be
independent, simulations are focused predominately on calculating the
response of a single {\arianna} station. For this purpose, the
interaction volume is chosen to be a rectangular prism having a height
575~{\m}, roughly equal to the ice thickness (see \sect{intro}), and a
horizontal cross section ranging from 3 to 10~$\km^2$, depending on
the neutrino energy.  Consistent ice depth measurements at multiple
locations, separated by 1~{\km}, have been performed at the {\arianna}
site~\cite{AriannaIcePaper}. These measurements have motivated the use
of a simple uniform ice thickness model in the simulations. The
neutrino interaction vertices are uniformly distributed within the
fiducial volume and the neutrino arrival directions are isotropically
distributed.

The neutrino energy is selected randomly according to a specified
flux. For the results presented in this article, the ESS cosmogenic
flux~\cite{Engel:2001hd} described in
\sect{arianna:performance:nurates} has been used. Neutrinos of each
flavor have been simulated with equal arrival rates, consistent with
the 1:1:1 flavor ratio expected of neutrinos generated by pion decay
at distant sources~\cite{Gaisser:1994yf}. Roughly two-thirds of the
neutrinos undergo a charged-current interaction, while the remainder
interact via the neutral current. The fraction of the neutrino energy
carried over to the resulting hadronic shower, the inelasticity $y$,
is randomly chosen following the distributions presented in
Ref.~\cite{Gandhi:1995tf}. This leads to an average inelasticity of
20\% for cosmogenic neutrinos in the energy range of interest to
{\arianna}.

The maximum strength of the electric field is
parametrized~\cite{AlvarezMuniz:2000fw} on the {\cheren} cone 1~{\m}
from the neutrino interaction vertex. This field strength is
proportional to the fraction of neutrino energy deposited into the
hadronic shower. The strength of the electric field at angles
off the {\cheren} cone, $\delta\theta_c$, is parametrized by a
Gaussian whose width is calculated according to
Ref.~\cite{AlvarezMuniz:1998px} for hadronic and
Ref.~\cite{AlvarezMuniz:1997sh} for electromagnetic showers.

For charged current $\nu_{e}$ interactions, the field strength is
increased by radiation from the prompt electromagnetic shower. For
such interactions at energies above $\sim\!10^{9}$~{\gev}, however,
the longitudinal shower profile increases faster than
$\log{E_{\nu_{e}}}$, rising as fast as
$\sqrt{E_{\nu_{e}}}$~\cite{Gerhardt:2010bj}. These elongated showers
lead to a sharper reduction of the electric field strength at angles
off the {\cheren} cone. This Landau-Pomeranchuk-Migdal (LPM) effect is
simulated by reducing the Gaussian width of the {\cheren} cone spread
by an amount proportional to $E_{\nu_{e}}^{-1/3}$.

The electric field is then propagated from the neutrino interaction
vertex to the detector. The ice is modeled by a 75~{\m} firn layer
atop a 500~{\m} ice shelf. The bulk ice is taken to have a refractive
index $n=1.78$~\cite{Radioglaciology}. The index of refraction in the
firn layer is calculated as a function of depth using the Schytt
model~\cite{schytt} combined with previous ice core measurements taken
from the Ross Ice Shelf~\cite{Barrella:2010vs}, varying from $n=1.30$
at the surface to $n=1.78$ in the ice.
%% \begin{equation}
%%    \label{arianna:eq:iceidxref}
%% n(z) = 1.0 + 0.86(1.0 - 0.638e^{-z/34.7{\rm m}})
%% \end{equation}
%% \noindent%
%% where $n$ is the index of refraction in the ice as a function of
%% depth, $z$.

Radio signals propagating through the ice are attenuated by a factor
that depends only on column depth. The frequency dependence of the
attenuation length is averaged over the bandwidth of the LPDA, taken
to be around 100~{\mhz} to 1~{\ghz}~\cite{AriannaTemplatePaper}. The
variation of attenuation length with depth arises from the changing
temperature of ice with depth, ranging from roughly -30{\dg}~C at the
surface of the ice to -2{\dg}~C at bottom of the ice shelf. This depth
variation results in an average attenuation length of
$400\pm18~{\m}$. Further power may be lost for radio pulses reflecting
off the ice-seawater boundary. The simulations halve the power of the
pulse (-3~{\db}) upon reflection, although in situ measurements show
the reflectivity, $R$, of the boundary to be between
$\sqrt{R}=0.82\pm0.07$~\cite{AriannaIcePaper}.

\subsubsection{Time Domain}
\label{arianna:sims:time}

The time domain response of the detector has been studied by
constructing a collection of ``waveform templates'' that quantify the
voltage measured by an antenna over time for an Askaryan radio
pulse. Each waveform template is calculated for a radio frequency (RF)
pulse arriving at a particular angle with respect to the antenna, as
well as for a particular observation angle $\delta\theta_c$ relative
to the {\cheren} cone.

%%  The measured responses of the {\arianna} LPDA
%% antenna and amplifier~\cite{AriannaNIMPaper} are used to calculate the
%% time-dependent waveform that would be recorded given the incident
%% radio pulse. This detector simulation is also used to predict the
%% response of {\arianna} to cosmic ray showers, as simulated by the
%% CoREAS package~\cite{Huege:2013vt, CoReasICRC2013, Huege:2013yra}.

The electric field produced by the Askaryan effect in ice is
calculated as a function of time~\cite{AlvarezMuniz:2010ty,
  AlvarezMuniz:2011ya} for an electromagnetic particle shower of a
specified energy. The charge excess is modeled as a pancake
with a charge profile that varies as the shower propagates. The effect
of the lateral structure of the particle shower on the electric field
is parametrized as a function of time using a form factor obtained
from shower simulations~\cite{AlvarezMuniz:2000fw}.

The measured responses of the {\arianna} amplifier and LPDA are
convolved with the electric field in order to produce the voltage
observed by an antenna as a function of time. The propagation of the
electric field through the ice reduces the strength of the field, but
has a negligible effect on the relative frequency content compared to
the impact of the hardware response. The antenna response is
quantified in bins of two angles, both relative to boresight: one in
the plane of the tines (the E-plane) and one normal to the plane of
the tines (the H-plane).

Full details of the procedure used to calculate the time-dependent
waveform templates may be found in Ref.~\cite{AriannaTemplatePaper}.

\subsection{Expected Performance}
\label{arianna:performance}

The properties of neutrino interactions that produce a radio pulse at
a station with sufficient power to trigger the detector have been
studied. Studies of the angular and energy resolution of {\arianna}
are performed using a simulation of an eight downward LPDA station
configuration. Simulated stations are triggered when the observed
radio signal is larger than four times the noise on at least three out
of the eight downward facing antennas. This threshold level has been
achieved in situ with stations running a two out of four antenna
channel trigger (see \sect{hra:data}).

\subsubsection{{\arianna} Aperture}
\label{arianna:performance:aperature}

\begin{figure}[t!]
   \begin{center}
     \includegraphics[width=\linewidth]{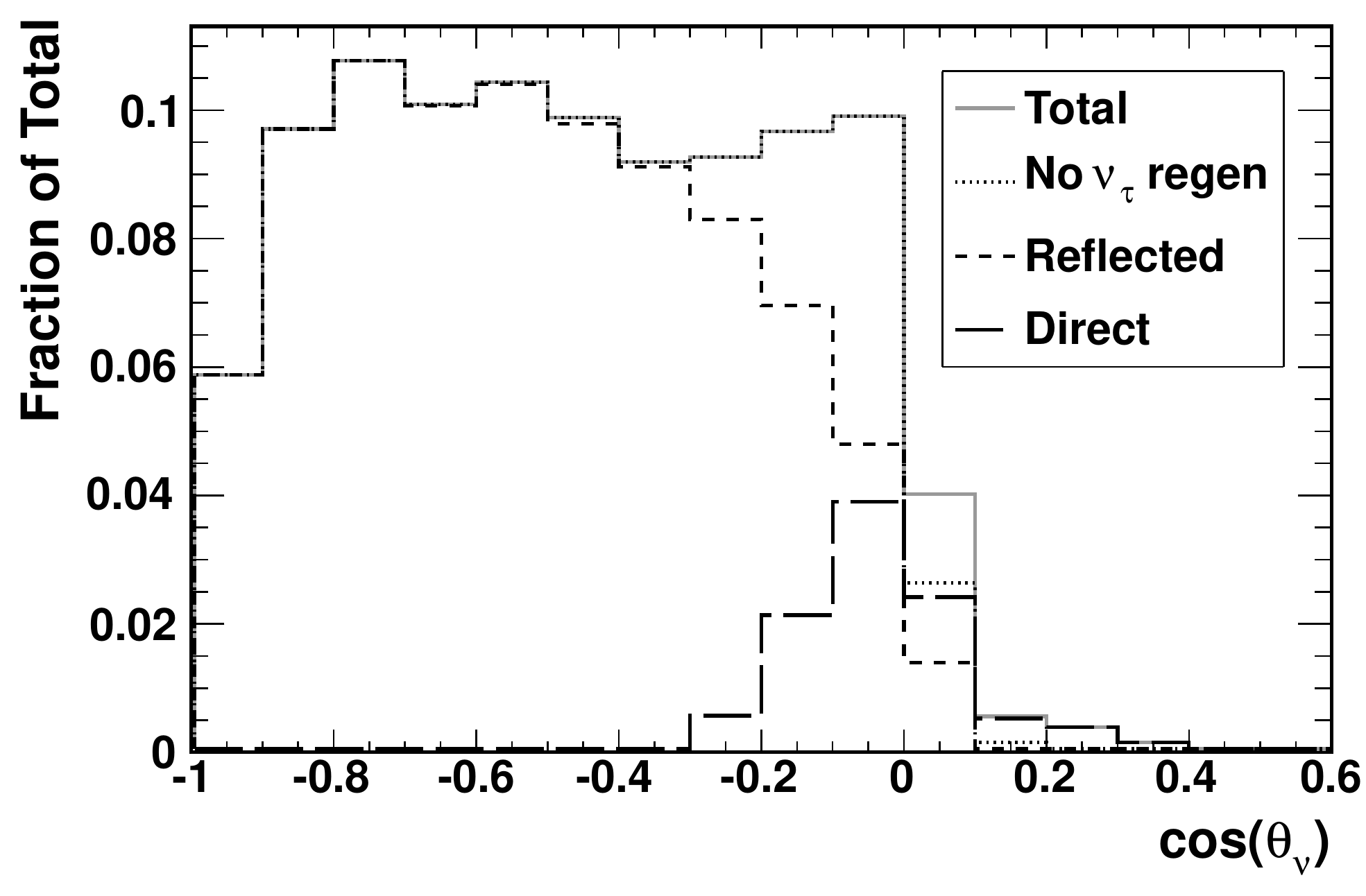}
   \end{center}
   \caption{\label{arianna:fig:angsensOnly}
     The number of neutrino triggers observed in each bin relative to
     the total number of cosmogenic~\cite{Engel:2001hd} neutrino
     triggers as a function of the local zenith neutrino arrival
     angle. The station triggers on neutrinos arriving from the
     horizon through either direct or reflected radio pulses. The
     majority of triggers are due to reflected pulses from locally
     down-going neutrinos. Figure adapted from
     Ref.~\cite{KamleshPhd}.}
\end{figure}

\begin{figure}[t!]
   \begin{center}
     \includegraphics[width=\linewidth]{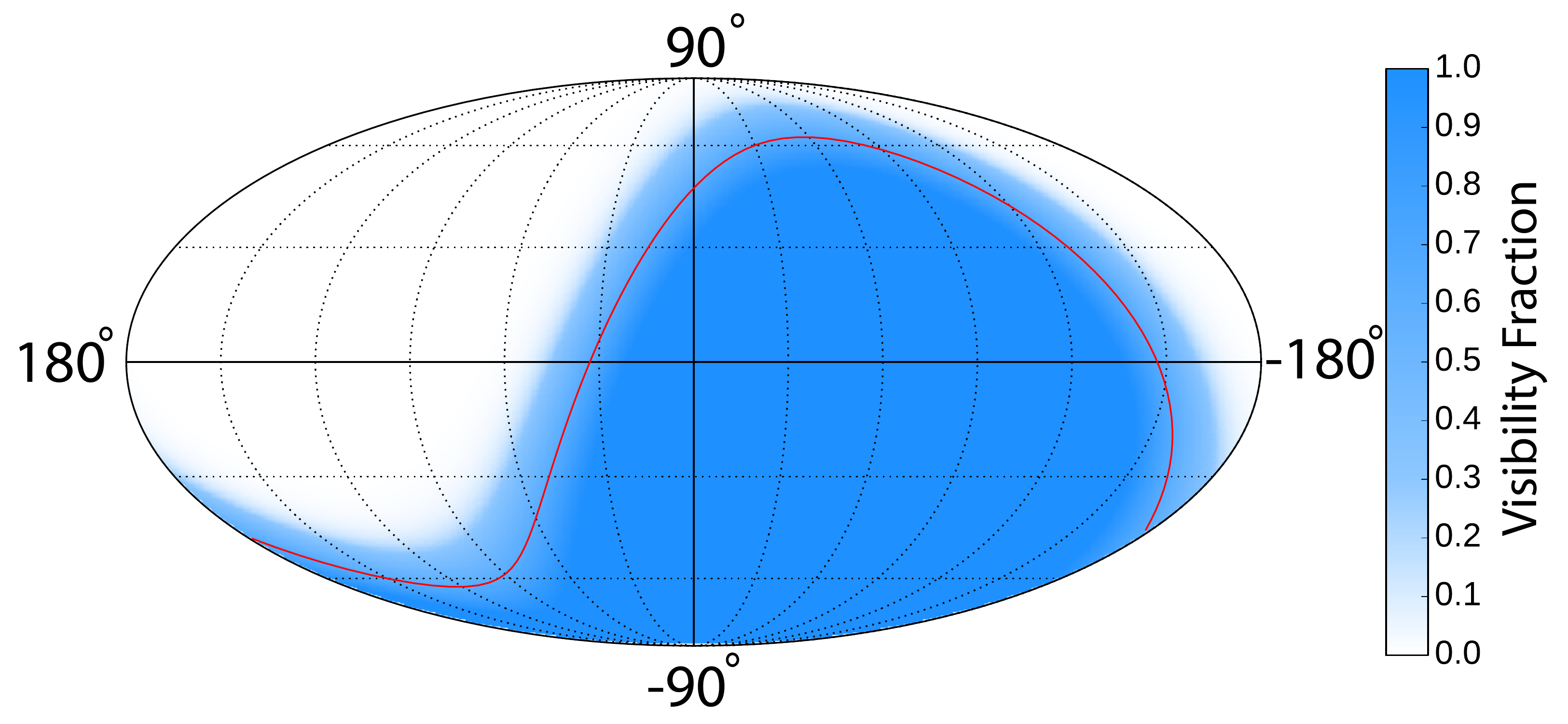}
   \end{center}
   \caption{\label{arianna:fig:skymap}
     The angular coverage in Galactic coordinates. The color scale
     represents the fraction of livetime that a patch of the sky is
     visible. The solid line shows the sky visible to an
     {\arianna}-like detector at the South Pole.}
\end{figure}

Neutrinos that trigger an {\arianna} station arrive predominately from
the sky above the station, creating a radio pulse that reflects off
the ice and water interface at the bottom of the ice
shelf. \Fig{arianna:fig:angsensOnly} shows the relative sensitivity of
an {\arianna} station to neutrinos, averaged over flavor, as a function
of local zenith angle. In addition to down-going neutrinos, {\arianna} is
also sensitive to neutrinos arriving from the horizon. Such events may
be triggered either through a reflected pulse or by a direct
observation of the {\cheren} wavefront, depending on geometry.

The portion of the sky that {\arianna} observes with this angular
sensitivity is shown in Galactic coordinates in
\fig{arianna:fig:skymap}. The line in the figure represents the view
of an {\arianna}-like detector located at the South Pole. Note that
the local zenith acceptance of an ARA-like detector falls off rapidly
for neutrinos originating more than 45{\dg} above the
detector~\cite{Allison:2011wk}. Thus, an ARA-like detector would have
reduced visibility of the sky around the southern polar region.

\begin{figure}[t!]
   \begin{center}
     \includegraphics[width=\linewidth]{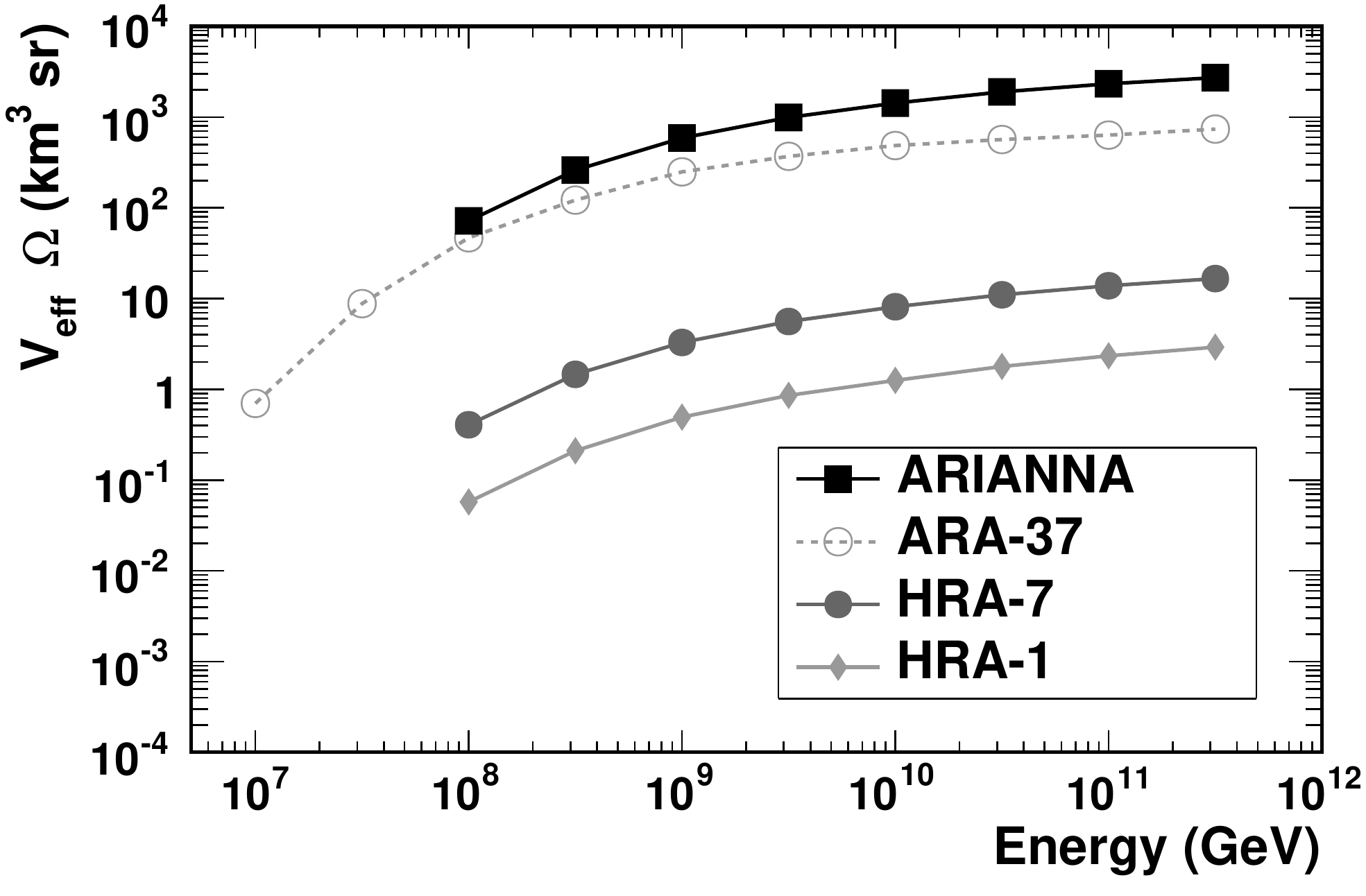}
   \end{center}
   \caption{\label{arianna:fig:effvol}
     The effective volume times total viewing angle, at the trigger
     level with $4\sigma$ thresholds, of a 1296 station {\arianna}
     telescope as a function of energy. The effective volumes are
     averaged over neutrino flavors as well as over neutrinos and
     anti-neutrinos. Also shown are the $\veff \Omega$ for a single
     HRA station, a seven station HRA, and the ARA-37
     detector~\cite{Allison:2011wk}.}
\end{figure}

\begin{figure}[t!]
   \begin{center}
     \includegraphics[width=\linewidth]{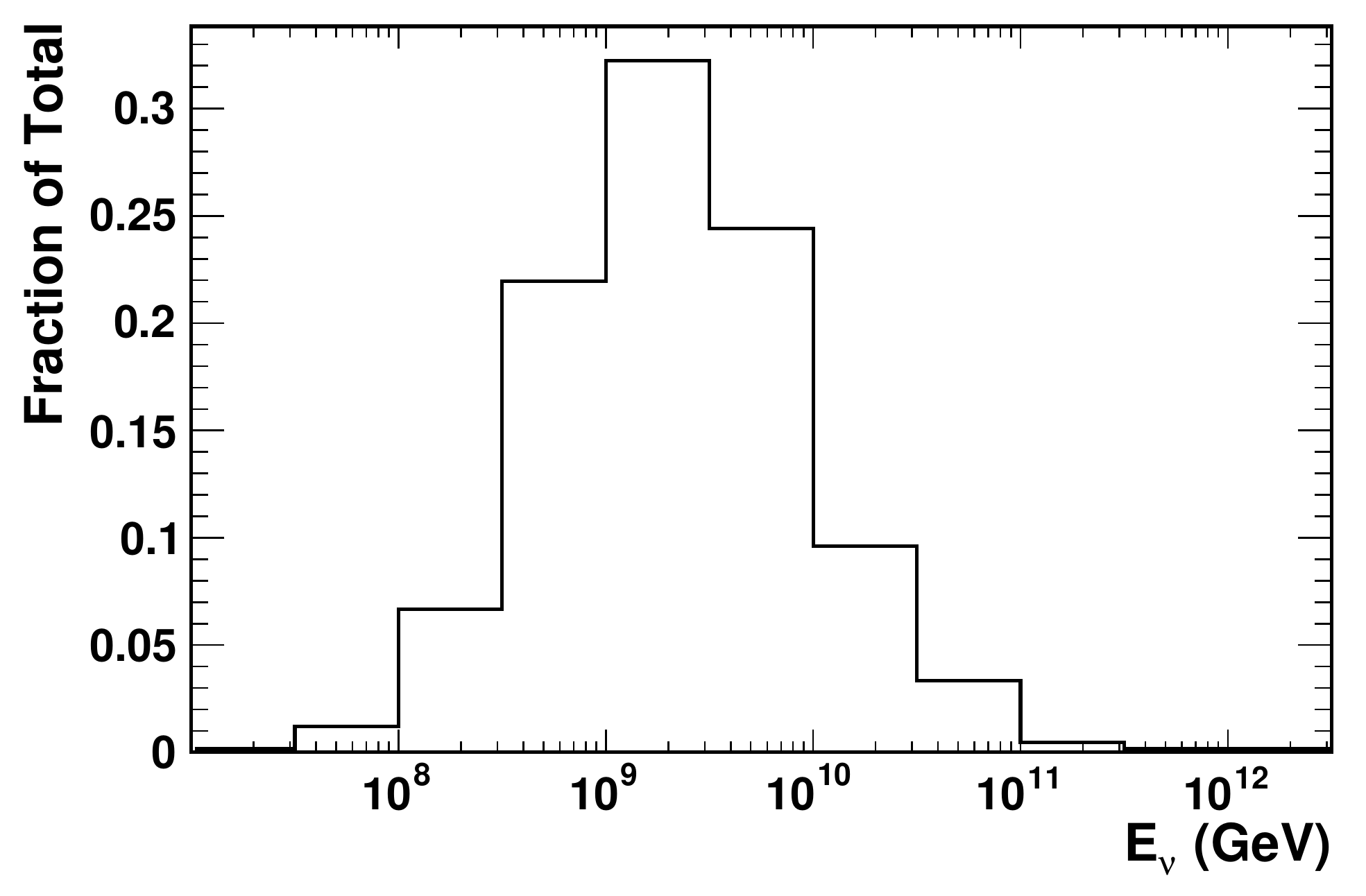}
   \end{center}
   \caption{\label{arianna:fig:energysens}
     The number of neutrino triggers observed in each bin relative to
     the total number of cosmogenic~\cite{Engel:2001hd} neutrino
     triggers as a function of the true simulated neutrino
     energy. Figure adapted from Ref.~\cite{KamleshPhd}.}
\end{figure}

The effective volume of the full {\arianna} telescope in which
neutrino interactions will be triggered has been studied as a function
of energy, presented in \fig{arianna:fig:effvol}, and is compared to
that of the ARA-37 experiment~\cite{Allison:2011wk}. The effective
volume shown in \fig{arianna:fig:effvol} is integrated over the viewing
angle of {\arianna}, allowing the expected number of neutrino triggers
due to a neutrino flux, $\Phi$, to be calculated as
\begin{equation}
   \label{arianna:eq:ntrgEffvol}
d\!N(E) = \Phi(E)\,\frac{\varepsilon \veff(E)\,\Omega}{\lint(E)}\,\tlive\,d\!E
\end{equation}
\noindent%
where $d\!N$ is the number of neutrinos in an energy bin, $E$ is the
average neutrino energy in the bin, $d\!E$ is the width of the bin,
$\varepsilon$ is the efficiency with which neutrino triggers are
preserved by an analysis, $\veff$ is the effective volume at the
trigger level, $\Omega$ is the viewing angle in steradians, $\lint$ is
the water-equivalent interaction length of neutrinos in the ice
($\approx\!10^3~${\km} at $10^{9}~${\gev}) and $\tlive$ is the
livetime of the experiment.

\begin{figure}[t!]
   \begin{center}
     \includegraphics[width=\linewidth]{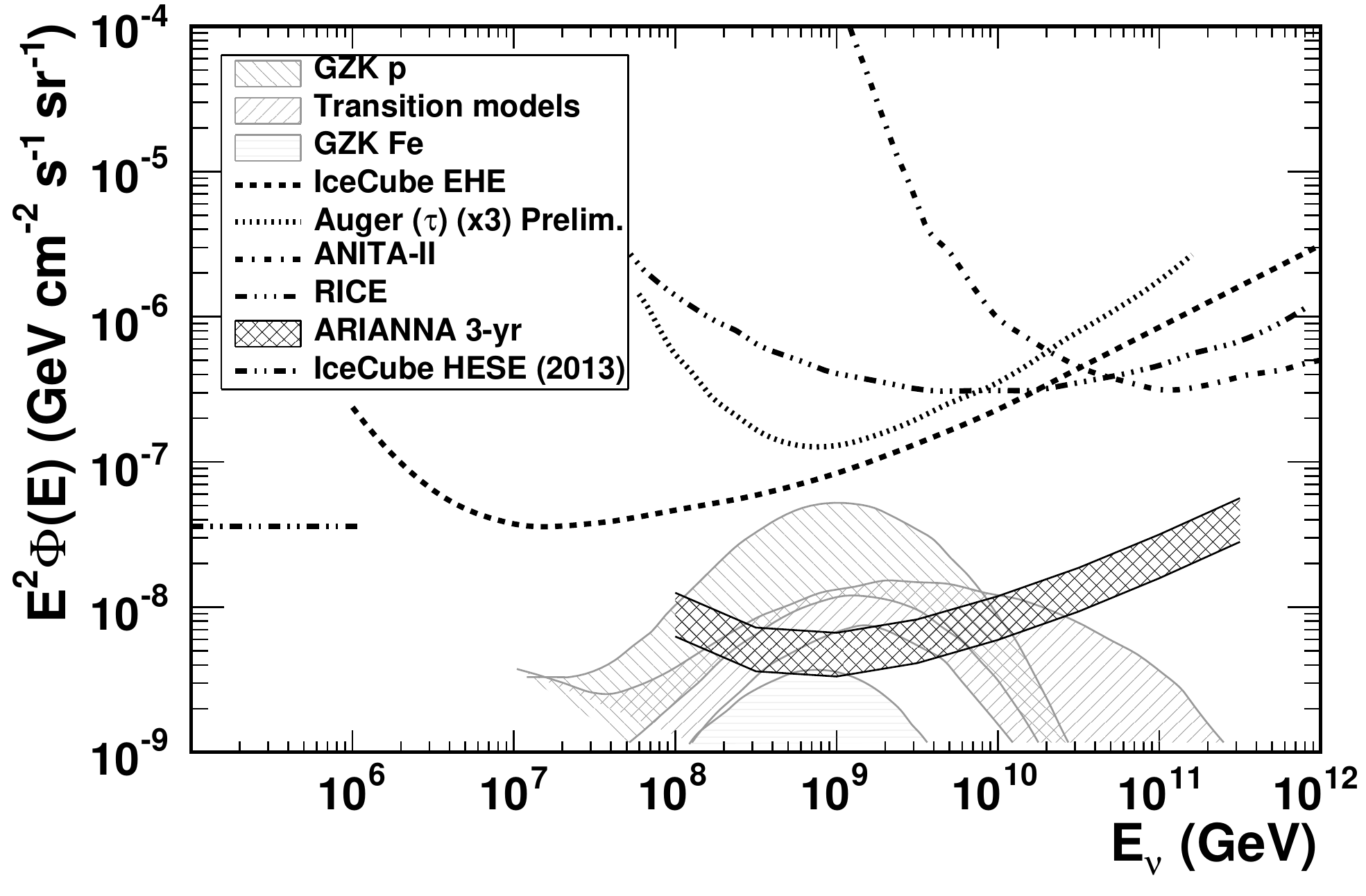}
   \end{center}
   \caption{\label{arianna:fig:fluxlims}
     The all flavor ${\nu+\bar{\nu}}$ differential sensitivity of a
     1296 station {\arianna} telescope running for 58\% of three years
     with a signal efficiency of 83\% (see \sect{ana:extrap}). The
     sensitivity is calculated in a sliding decade-wide neutrino
     energy bin. See text for a discussion of the width of the
     sensitivity band. Limits on the flux of neutrinos are shown for
     several experiments~\cite{Abreu:2012zz, Abbasi:2011ji,
       Kravchenko:2011im, Aartsen:2013jdh}, with
     ANITA~\cite{PhysRevD.82.022004,Gorham:2010kv} providing the most
     stringent limits at the highest energies and
     IceCube~\cite{Aartsen:2013dsm} at lower energies. The cosmogenic
     neutrino flux predicted by several models is shown for different
     assumptions of the cosmic ray composition~\cite{Kampert:2012mx}.}
\end{figure}

This effective volume leads to an energy distribution of cosmogenic
neutrino triggers that has 90\% of detected neutrinos between
$10^{8.4}~${\gev} and $10^{10.4}~${\gev}, as shown in
\fig{arianna:fig:energysens}. On an absolute scale, the all flavor
${\nu+\bar{\nu}}$ sensitivity of {\arianna} to trigger on neutrinos is
presented in \fig{arianna:fig:fluxlims}. This sensitivity has been
calculated for 1296 {\arianna} stations running for three calendar
years, with a livetime equal to 58\% of each calendar year, and a
signal efficiency of 83\% (see \sect{ana:extrap}). The fractional
livetime corresponds to stations powered by batteries and/or solar
panels and is based on the observed behavior of previously and
currently deployed prototype stations. The sensitivity is calculated
as the average Neyman upper limit with an expected Poissonian
background of 0.3~events (see \sect{ana:extrap}); i.e.~$d\!N=2.47$
neutrinos in the sliding decade-wide energy bin.

The systematic uncertainty band on the sensitivity shown in
\fig{arianna:fig:fluxlims} accounts for uncertainties on the models
used to describe various physics processes in the simulations. The
neutrino-nucleon cross section is calculated from parton distribution
functions extrapolated to the as yet unmeasured low-$x$ values
appropriate for $E_{\nu}{\ge}10^{7}$~{\gev}, an estimation that may
over or underestimate rates. The simulation of the LPM effect does not
fragment the charge excess into separate clumps, which would give rise
to competing effects: a reduction of the overall RF signal strength
and an increase of the relative amplitude at angles away from the
{\cheren} cone.

Other models employed by the simulations are understood to either
increase or decrease expected rates. No RF contribution of the $\mu$
lepton arising from charged current $\nu_{\mu}$ interactions is
simulated, which underestimates trigger
rates~\cite{Wang:2013njo}. However, the $\nu_{\tau}$ trigger rates are
likely overestimated since the $\tau$ lepton resulting from a
$\nu_{\tau}$ interaction is not propagated. Instead, the shower with
the greater energy, either from the $\nu_{\tau}$ interaction or the
$\tau$ decay (if it produces a shower), is simulated at the neutrino
interaction location. Thus, $\tau$ lepton decays that produce very
high energy showers outside the fiducial volume are allowed to trigger
a station.

Uncertainties on the ice properties used in the simulations also
contribute to the sensitivity band. Simulated pulses lose half their
power upon reflection, however measurements at the site suggest a
significantly larger reflection coefficient. Similarly, ice property
measurements at the site suggest the RF signal loss due to refraction
and attenuation in the firn layer is likely overestimated.

\subsubsection{Expected Neutrino Rates}
\label{arianna:performance:nurates}

\begin{table*}[t!]
   \begin{center}
      \begin{tabular}{|l|l|l|l|}
\hline
\hline
Neutrino Model & Model Type &
\multicolumn{2}{c|}{$N_{\nu}$ Triggers ($E_{\nu}\!>\!10^{8}~{\gev}$)}\\
  &  & {\arianna} %& ARA~\cite{Allison:2011wk}
 & IceCube~\cite{Aartsen:2013dsm}\\
\hline
%ESS (2001) Fig.~4~\cite{Engel:2001hd} & No source evolution & 21 &  & \\
%\hline
ESS (2001)~\cite{Engel:2001hd} &  $m\!=\!4$, $\Omega_{M}\!=\!1$ & 55 & \\
\hline
%WB (1999)~\cite{Waxman:1998yy} & $E_{\nu}^{-2}$ No source evolution & 18 & 27 & \\
%\hline
WB (1999)~\cite{Waxman:1998yy} & $E_{\nu}^{-2}$ QSO source evolution & 65 & \\
\hline
Yuksel {\etal} (2007)~\cite{Yuksel:2006qb} & $E_{\nu}^{-2}$ GRB source evolution & 100 & \\
%Yuksel {\etal} (2007)~\cite{Yuksel:2006qb} & $E_{\nu}^{-1.5}$ GRB source evolution & 241 & \\
\hline
Kotera {\etal} (2010)~\cite{Kotera:2010yn} & Protons, SFR1 evolution & 7.3 & 0.46 (0.64) \\
\hline
Kotera {\etal} (2010)~\cite{Kotera:2010yn} & Protons, GRB2 evolution & 9.0 & 0.48 (0.67) \\
\hline
Kotera {\etal} (2010)~\cite{Kotera:2010yn} & Protons, FRII evolution & 48 & 2.9 (4.0) \\
\hline
Yoshida {\etal} (1993)~\cite{Yoshida:1993pt} & $m\!=\!4$, $z_{\mathit{max}}\!=\!4$ & 34 & 2.0 (2.8) \\
\hline
Ahlers {\etal} (2010)~\cite{Ahlers:2010fw} & $\emin\!=\!10^{10}$~{\gev} (best fit) & 26 & 1.5 (2.1) \\
\hline
Ahlers {\etal} (2010)~\cite{Ahlers:2010fw} & $\emin\!=\!10^{10}$~{\gev} (maximal) & 58 & 3.1 (4.3) \\
\hline
Kotera {\etal} (2010)~\cite{Kotera:2010yn} & Mixed composition & 7.4 & \\
\hline
Kotera {\etal} (2010)~\cite{Kotera:2010yn} & Pure Iron & 2.5 &  \\
\hline
Ave {\etal} (2005)~\cite{Ave:2004uj} & Pure Iron, $m\!=\!4$, $z_{\mathit{max}}\!=\!1.9$ & 18 &  \\
\hline
Olinto {\etal} (2011)~\cite{Olinto:2011ng} & Pure Iron, $\emax/Z\!=\!10^{11}$~{\gev} & 0.097 & \\
\hline
Aartsen {\etal} (2014)~\cite{Aartsen:2014gkd} & $E_{\nu}^{-2.3}$ IceCube best fit & 2.8 & \\
\hline
Fang {\etal} (2013)~\cite{Fang:2013vla} & Young pulsar sources & 43 & \\
\hline 
\hline
      \end{tabular}
   \end{center}
   \caption{   \label{arianna:tab:models}
     The expected number of triggers due to neutrinos of all flavors
     with $E_{\nu}\!>\!10^{8}~{\gev}$ in 1296 {\arianna} stations
     after running for 3 calendar years given different models of the
     cosmogenic neutrino flux. A flavor ratio of 1:1:1 at Earth is
     assumed. A realistic livetime of 58\% per year and a signal
     efficiency of 83\% (see \sect{ana:extrap}) has been used for the
     {\arianna} rates. See text for an explanation of the model
     types. For reference, if 0.3~background events are expected in
     the data set (see \sect{ana:extrap}), 6.4 neutrino events would
     push the number of observed events beyond a $5\sigma$ background
     fluctuation in 50\% of experiments (prior to any trial factor
     penalties). Published neutrino rates for IceCube with 333.5 days
     of IC40, 285.8 days of IC79 and 330.1 days of IC86
     data~\cite{Aartsen:2013dsm} are shown where available. The
     numbers in parentheses show the IceCube rates increased by 39\%
     to facilitate direct comparison with the {\arianna} 3 year
     rates.}
\end{table*}

The number of neutrino triggers that {\arianna} can expect to record
after 3~calendar years of running is presented in
\tab{arianna:tab:models} for a variety of neutrino flux models. Unless
otherwise stated, neutrino fluxes are generated through photopion
production by a flux of cosmic rays. The evolution of cosmic ray
sources follows the evolution of potential source populations, such as
the star formation history. This is typically approximated by an
increasing emissivity per co-moving volume proportional to $(1+z)^{m}$
out to a first break point at $z\!=\!z_{\mathit{max}}$, where $z$ is
the redshift of the source.

The ESS model~\cite{Engel:2001hd} shown in \tab{arianna:tab:models}
assumes cosmic sources evolve with $m\!=\!4$ and
$z_{\mathit{max}}\!=\!1.9$ in a flat universe. The WB
model~\cite{Waxman:1998yy} follows the luminosity density evolution of
quasi-stellar objects (QSOs) with $m\!=\!3$ for
$z_{\mathit{max}}\!=\!1.9$. The Yuksel {\etal}~\cite{Yuksel:2006qb}
model parameterizes a very strong source evolution according to the
gamma ray-burst (GRB) rate, with $m\!=\!4.8$ up to
$z_{\mathit{max}}\!=\!1$.  Three of the Kotera
{\etal}~\cite{Kotera:2010yn} models vary the source evolution of a
flux of protons, with $m\!=\!3.4$ up to $z_{\mathit{max}}\!=\!1$ for a
star formation rate (SFR1) evolution. The GRB2 model from Kotera
{\etal} closely follows the SFR1 evolution, but continues to gradually
increase beyond $z\!>4$. The Kotera {\etal} FRII model employs a very
steep evolution out to $z_{\mathit{max}}\!=\!4$. The Yoshida and
Teshima~\cite{Yoshida:1993pt} also assumes a strong evolution,
$m\!=\!4$, out to $z_{\mathit{max}}\!=\!4$. The Ahlers
{\etal}~\cite{Ahlers:2010fw} models take $m\!=\!4.6$ until
$z_{\mathit{max}}\!=\!2$ and constrain the energy spectra of particles
injected to the cosmic ray accelerating source using gamma ray
measurements taken by the Fermi-LAT. The cross-over energy between
galactic and extragalactic cosmic rays is parameterized by a lower
energy break in the injection spectra of $\emin\!=\!10^{10}$~{\gev}.

The possibility of a chemical composition of the cosmic ray flux that
is not purely proton is also explored. The Kotera {\etal} mixed
composition model assumes that the chemical composition of particles
injected to the cosmic ray source matches the composition of the low
energy galactic cosmic ray flux. The Kotera {\etal} pure iron model
tests the case of a purely iron nucleus composition. In both cases,
the sources evolve according to the SFR1 parameterization. The Ave
{\etal}~\cite{Ave:2004uj} model also uses a purely iron composition,
but includes a stronger source evolution of $m\!=\!4$ out to
$z_{\mathit{max}}\!=\!1.9$. The Olinto {\etal}~~\cite{Olinto:2011ng}
tests a purely iron composition combined with a uniform source
evolution and a cutoff in the source acceleration at
$\emax/Z\!=\!10^{11}$~{\gev}.

The Aartsen {\etal}~\cite{Aartsen:2014gkd} flux represents the best
fit to the neutrino flux measurement obtained by the IceCube
collaboration after the observation of 37~neutrino candidate events
with energies up to 2~{\pev}. A measurement of this flux by {\arianna}
would extend the currently measured spectrum to higher neutrino
energies by at least two orders of magnitude.

{\arianna} can also search for alternative sources of UHE neutrinos
such as young pulsars that accelerate particles through surrounding
supernova remnant material, as modeled by Fang
{\etal}~\cite{Fang:2013vla}. In this model, sources evolve according
to the star formation rate.

{\arianna} can expect to see about 13 times as many neutrinos above
$10^{8}~{\gev}$ as IceCube, depending on the flux model, and a
comparable number of neutrinos to ARA~\cite{Allison:2011wk} for
similar model parameters. This increase in sensitivity relative to
current limits will create the opportunity to study nearly all mixed
composition models (currently favored by Auger data) that include a
power law injection spectrum, and to probe alternate scenarios of
cosmic ray acceleration such as young pulsar sources.

\subsubsection{Angular Resolution}
\label{arianna:performance:angres}

The angular resolution of an {\arianna} station has been estimated
through the use of a simple reconstruction procedure on simulated
neutrino events. The arrival direction of the radio pulse is found by
fitting the time difference between pulses in different antennas, on a
single {\arianna} station, to a planar wavefront. Thanks to the better
than 100~{\ps} timing of {\arianna} stations~\cite{AriannaNIMPaper},
the angular error on the direction of the RF signal at the station is
better than 1{\dg}~\cite{ReedIcrc2013}.

To determine the direction of the neutrino, however, it is necessary
to estimate both the signal propagation direction as well as the
signal polarization direction. For example, a signal arriving
vertically upward and on the {\cheren} cone would be produced by a
neutrino with a zenith angle equal to the {\cheren} angle. Any azimuth
angle would produce the same upward propagating signal. The
polarization angle serves to break this degeneracy, as the
polarization is perpendicular to the signal propagation direction and
points away from the shower axis.

%% \begin{figure}[t!]
%%    \begin{center}
%%      \includegraphics[width=\linewidth]{img/Hra2013eightLpdaGeo}
%%    \end{center}
%%    \caption{\label{arianna:fig:eightLpdaGeo}
%%      The orientation of the eight LPDAs placed symmetrically around
%%      the DAQ box at a distance of 2.61~{\m}. The eight antenna station
%%      has been used to study the angular and energy resolution of
%%      {\arianna}.
%%      %\comm{Get better graphic.}
%%    }
%% \end{figure}

An {\arianna} station with eight downward-facing LPDAs has been
simulated in order to quantify the neutrino angular resolution. The
polarization of the incoming radio pulse is determined by the relative
amplitude of the pulses recorded by non-parallel antennas. The voltage
recorded by an antenna is proportional to the component of the
electric field vector in the direction parallel with the tines of the
LPDA. The constant of proportionality depends on the incoming
direction of the radio signal and on the in-ice antenna response.

%% In the frequency domain, this is expressed as
%% \begin{equation}
%%    \label{arianna:eq:voltfromE}
%% V^i(\nu) = \vec{\varepsilon}(\nu)\cdot\vec{p}^i
%% \left[\frac{1}{2\sqrt{2}} h_{eff}(\nu) \exp\left(-2 \ln2 
%%   \left( \frac{\theta^i_{inc}}{\theta_{h}(\nu)} \right)^2 \right) \right]
%% \end{equation}
%% \noindent%
%% where $\nu$ is the frequency, $V^i$ is the voltage recorded on the
%% $i^{\textrm{th}}$ LPDA, $\vec{\varepsilon}\cdot\vec{p}^i$ is the
%% component of the electric field (at the antenna) in the E-plane of the
%% LPDA, $h_{eff}$ is the effective height of the LPDA, $\theta^i_{inc}$
%% is the incident angle of the signal on the antenna and $\theta_{h}$ is
%% the half-power beam width of the LPDA. The effective height of the
%% antenna parametrizes the response of an antenna in firn ice to a
%% signal of a particular frequency and is related to the gain of the
%% antenna. The half-power beam width describes the opening angle over
%% which the antenna radiation power is half its maximum or better, at a
%% particular frequency.

The polarization is reconstructed by first finding the antenna with
the largest recorded pulse. The voltage recorded by each of the two
adjacent antennas is then used to obtain the magnitude of the electric
field component that is parallel to the tines of that antenna. The
ratio of these electric field components gives the tangent of the
azimuth angle of the polarization vector (the component in the plane
of the ice surface). The reconstruction is improved by averaging the
signal of each antenna with its parallel counterpart antenna prior to
calculating the ratio. This improves the signal to noise of the
measurement of each electric field component and gives a more accurate
estimate of the azimuth angle of the polarization vector.

The zenith angle of the polarization vector is then constrained by the
requirement that the polarization and signal propagation vectors be
orthogonal. This yields two degenerate polarization vectors, each
having the same azimuth angle and each being perpendicular to the
signal propagation. Given that only the signal propagation vector is
needed to measure the energy of the neutrino, this degeneracy does not
affect a measurement of the diffuse neutrino flux. For point source
searches, however, it will result in two possible neutrino source
directions. The correct degenerate direction will always roughly point
back to the source, while the incorrect degenerate direction will vary
depending on the station orientation (and time). Thus, only a single
point of excess signal would be expected for a given neutrino source.

%% \begin{figure}[t]
%%    \begin{center}
%%      \begin{minipage}{0.49\linewidth}%
%%        \includegraphics[width=\linewidth]{img/Hra2013thetareso}
%%      \end{minipage}%
%%      \begin{minipage}{0.49\linewidth}%
%%        \includegraphics[width=\linewidth]{img/Hra2013phireso}
%%      \end{minipage}%
%%    \end{center}
%%    \caption{\label{arianna:fig:angreso}
%%      Angular difference between the reconstructed and true neutrino
%%      direction in the local (a)~zenith and (b)~azimuth directions is
%%      found to be $\sigma_{\theta}=2.9{\dg}$ and
%%      $\sigma_{\phi}=2.5{\dg}$, respectively, for an eight-antenna
%%      {\arianna} station.}
%% \end{figure}

\begin{figure}[t!]
   \begin{center}
     \includegraphics[width=\linewidth]{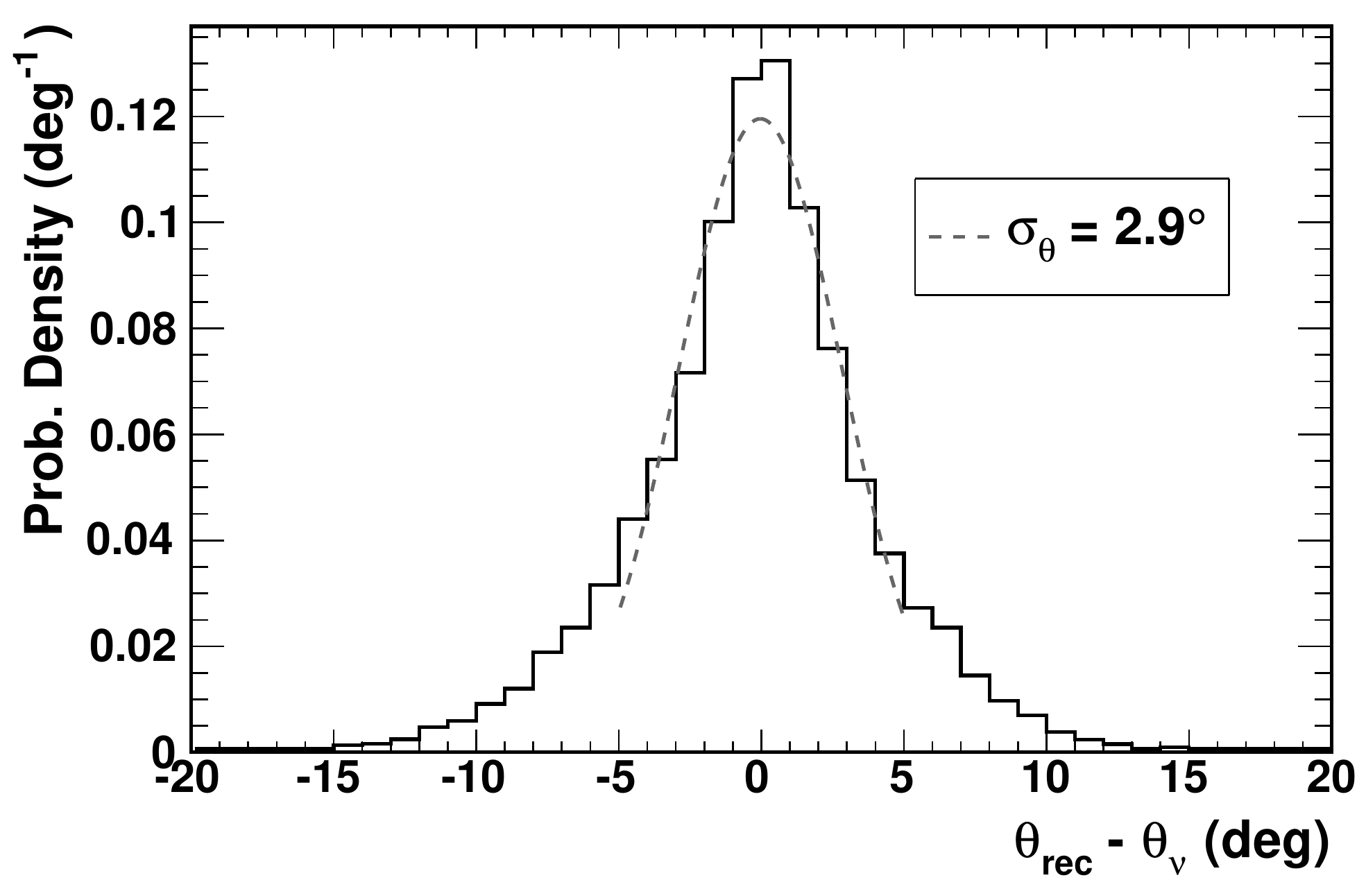}
   \end{center}
   \caption{\label{arianna:fig:thetareso}
     The angular difference between the reconstructed and true
     neutrino direction in the local zenith direction. A resolution of
     $\sigma_{\theta}=2.9{\dg}$ is found for an eight-antenna
     {\arianna} station. Figure adapted from Ref.~\cite{KamleshPhd}.}
\end{figure}

\begin{figure}[t!]
   \begin{center}
     \includegraphics[width=\linewidth]{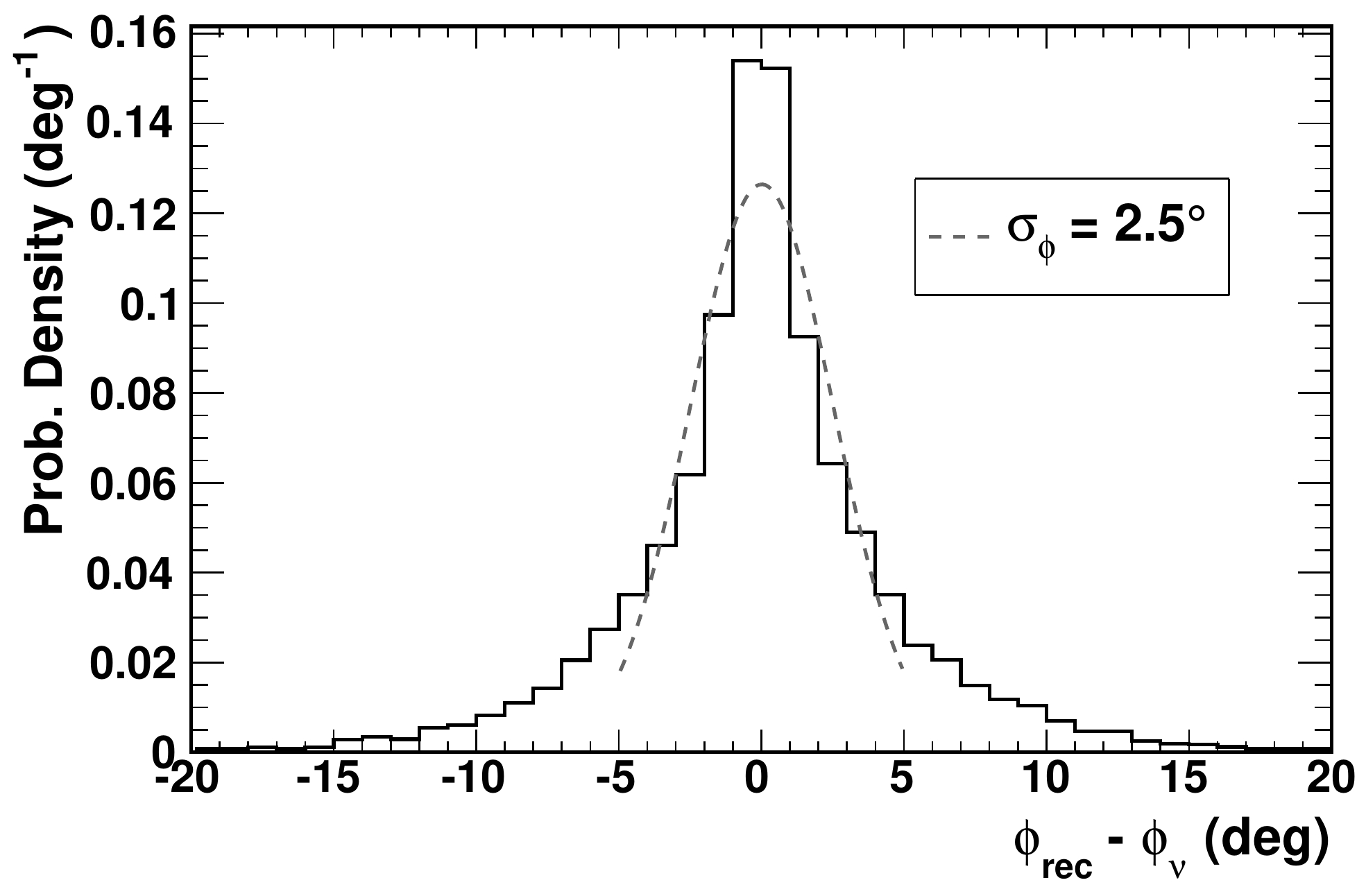}
   \end{center}
   \caption{\label{arianna:fig:phireso}
     The angular difference between the reconstructed and true
     neutrino direction in the local azimuth direction. A resolution
     of $\sigma_{\phi}=2.5{\dg}$ is found for an eight-antenna
     {\arianna} station.  Figure adapted from Ref.~\cite{KamleshPhd}.}
\end{figure}

The angular resolution expected for this method is shown in
\figs{arianna:fig:thetareso}{arianna:fig:phireso}. The resolution is
calculated using the more accurate (compared to the true direction) of
the two degenerate candidates, since this is the relevant quantity for
a neutrino point source search. The local zenith and azimuth angular
resolutions are $\sigma_{\theta}=2.9{\dg}$ and
$\sigma_{\phi}=2.5{\dg}$, respectively.

These values are conservative, in that they have been obtained by a
reconstruction that assumes that each antenna is observing the
electric field exactly at the {\cheren} cone,
i.e.~$\delta\theta_c=0$. In fact, on average $\delta\theta_c$ is found
to be about 2.2{\dg}, and some triggers are produced by antennas
observing signals up to 15{\dg} off-cone. The assumption that all
signals are on the {\cheren} cone is a major source of inaccuracy in
the reconstruction of the neutrino direction. Determining the
$\delta\theta_c$ to within 1.5{\dg} improves the neutrino direction
angular resolution by over 50\% in zenith and over 20\% in azimuth.

\subsubsection{Energy Resolution}
\label{arianna:performance:energyres}

The energy of an individual neutrino event is reconstructed by
determining the strength of the electric field at the station and
propagating it back through the ice to the shower location. The shower
energy is proportional to the amplitude of the electric field at the
shower. The constant of proportionality is calculated using measured
antenna response properties. The energy of the original neutrino is
then estimated using the average fraction of neutrino energy
transferred to a shower that results in a trigger. The trigger
requirement selects a subset of neutrino interactions for which a
large fraction of the energy, about 0.8, is transferred to the
shower. An average path length of radio pulses is used in lieu of a
shower vertex reconstruction. The average path length is determined
from simulations as a function of the zenith angle of the radio signal
propagation, which itself is readily determined using timing (see
\sect{arianna:performance:angres}).

There are two dominant sources of error on the neutrino energy
estimate. The first arises due to the Gaussian dependence of the
electric field amplitude on $\delta\theta_c$. Uncertainties on
$\delta\theta_c$ form the largest source of error on the energy
estimate. The unknown amount of energy transferred from the neutrino
to the charged particle shower is the second significant source of
error on the energy estimate. The distribution and average value of
these two parameters depend on neutrino flavor, which is assumed to be
unknown in the current analysis. As more sophistication is applied to
event reconstruction, it should be possible to identify the flavor and
thereby improve the energy resolution. Errors due to the inexact pulse
propagation length, electric field losses due to reflection at the ice
and water boundary and inaccuracies in the antenna response are each
on the order of 20-25\%, and are negligible by comparison.

\begin{figure}[t!]
   \begin{center}
     \includegraphics[width=\linewidth]{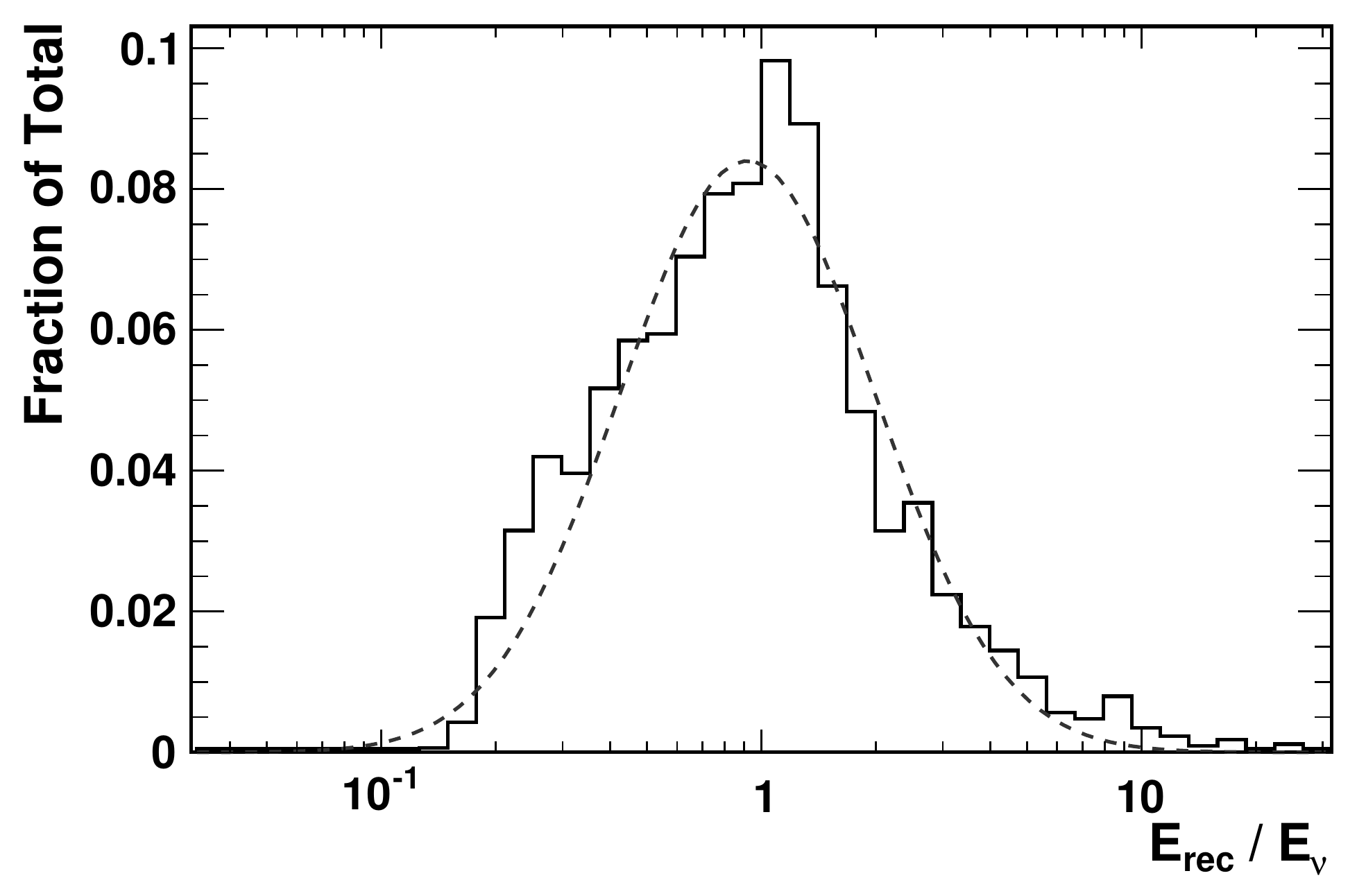}
   \end{center}
   \caption{\label{arianna:fig:energyreso}
     The energy resolution of an eight-antenna {\arianna} station to a
     cosmogenic flux~\cite{Engel:2001hd} is
     $\sigma(\log(\erec/E_\nu))\!=\!0.34$, so that
     $\sigma(\erec/E_\nu)\!=\!2.2$ for cosmogenic neutrinos, summed
     over neutrino flavor. Figure adapted from
     Ref.~\cite{KamleshPhd}.}
\end{figure}

In the analysis presented here, there is no attempt to determine the
angular deviation from the {\cheren} cone, $\delta\theta_c$.
Inserting the average value obtained from simulation studies, the
energy resolution is found to be $\sigma(\erec/E_\nu)\!\approx\!5$,
under the assumption that $\esh/E_\nu\!=\!0.8$ for triggered events,
where $\esh$ ($E_\nu$) is the energy of the shower
(neutrino). However, because the frequency and phase content of the
Askaryan pulse depends on $\delta\theta_c$, it can be exploited in
future analyses to reduce this uncertainty. Such a potential
measurement of $\delta\theta_c$ has been modeled in
Ref.~\cite{KamleshPhd} in order to investigate the reduction in energy
resolution. The result of this study is shown in
\fig{arianna:fig:energyreso}, which indicates that the average
neutrino energy resolution is reduced to
$\sigma(\erec/E_\nu)\!=\!2.2$, an improvement by a factor greater than
2.

%% \Fig{arianna:fig:energyreso} shows the cosmogenic neutrino energy
%% resolution obtained assuming $\esh/E_\nu=0.8$, where $\esh$ ($E_\nu$)
%% is the energy of the shower (neutrino), and assuming a resolution on
%% the {\cheren} observation angle of $\sim\!0.4{\dg}$ for signals
%% arriving near the cone. For signals measured off the {\cheren} cone,
%% the resolution is assumed to worsen as
%% $\sigma(\delta\theta_c)=0.3{\dg} + 0.1 \delta\theta_c$. This leads to
%% an average neutrino energy resolution of
%% $\sigma(\log(\erec/E_\nu))=2.2$ for a cosmogenic neutrino
%% flux~\cite{Engel:2001hd}, where $\erec$ is the reconstructed neutrino
%% energy. With no knowledge of the observation angle, the neutrino
%% energy resolution increases to $\sigma(\log(\erec/E_\nu)) \approx 5$.

%% components in the adjacent antennas as
%% \begin{equation}
%%    \label{arianna:eq:recoPolPhi}
%% \varepsilon_{\phi} = \arctan \left( 
%% \frac{\vec{\varepsilon}(\nu)\cdot\vec{p}^{m-1}}
%%      {\vec{\varepsilon}(\nu)\cdot\vec{p}^{m+1}} \right)
%% \end{equation}
%% \noindent%
%% where $m$ is the LPDA that recorded the largest pulse.

%% \subsubsection{From Eight to Four Antennas}
%% \label{arianna:performance:fourAntReso}

\subsection{Cosmic Ray Detection}
\label{arianna:cosmics}

Ultra-high energy cosmic rays are a plausible source of (reducible)
background for {\arianna}. Charged particle showers in the atmosphere
above the telescope, produced by UHE cosmic rays, can emit a
detectable radio pulse.  Much of this background is reduced naturally
by the 15~{\db} decrease in sensitivity of the LPDA back lobe relative
to boresight. The addition of two upward facing antennas eliminates
the remaining cosmic ray background. These antennas allow RF pulses
originating from the atmosphere to be efficiently distinguished from
neutrino pulses originating in the ice.

Dedicated cosmic ray simulations have been performed to study the rate
at which an {\arianna} station triggers on air showers and the
efficiency with which such events are separated from neutrino events.
The simulation of the cosmic ray air showers are performed using the
CoREAS software~\cite{Huege:2013vt, CoReasICRC2013,
  Huege:2013yra}. Proton interactions between $10^{8.4}$ and
$10^{10.5}~{\gev}$ are simulated by Corsika~\cite{Heck:1998vt} using
the QGSJetII-04~\cite{Ostapchenko:2005nj} hadronic model and weighted
by the high energy cosmic ray flux measured by the Auger
experiment~\cite{Abreu:2011pj}.

Cosmic rays are studied over a local zenith angle range from
0{\dg}-75{\dg} under the assumption of an isotropic flux. The azimuth
direction and interaction vertex position relative to the station are
varied for each combination of energy and zenith direction.

The radio pulse from the charged particle shower is generated and
propagated to the surface of the ice by CoREAS. The electric field at
the station in the frequency domain is convolved with the measured
{\arianna} antenna response, the result of which is then convolved
with the measured amplifier response to obtain the voltage expected on
each readout channel of the DAQ. A trigger is generated if the signal
in three or more of the eight downward facing antenna channels is
above four times the noise ($4\sigma$). This is the same trigger
criteria used in \sect{arianna:performance} to study neutrino
signals. Finally, finite bandwidth noise consistent with that observed
by HRA stations is added to the signal. The trigger is applied prior
to the addition of noise in order to avoid over-counting due to events
that will be rejected by any analysis.

\begin{figure}[t!]
   \begin{center}
     \includegraphics[width=\linewidth]{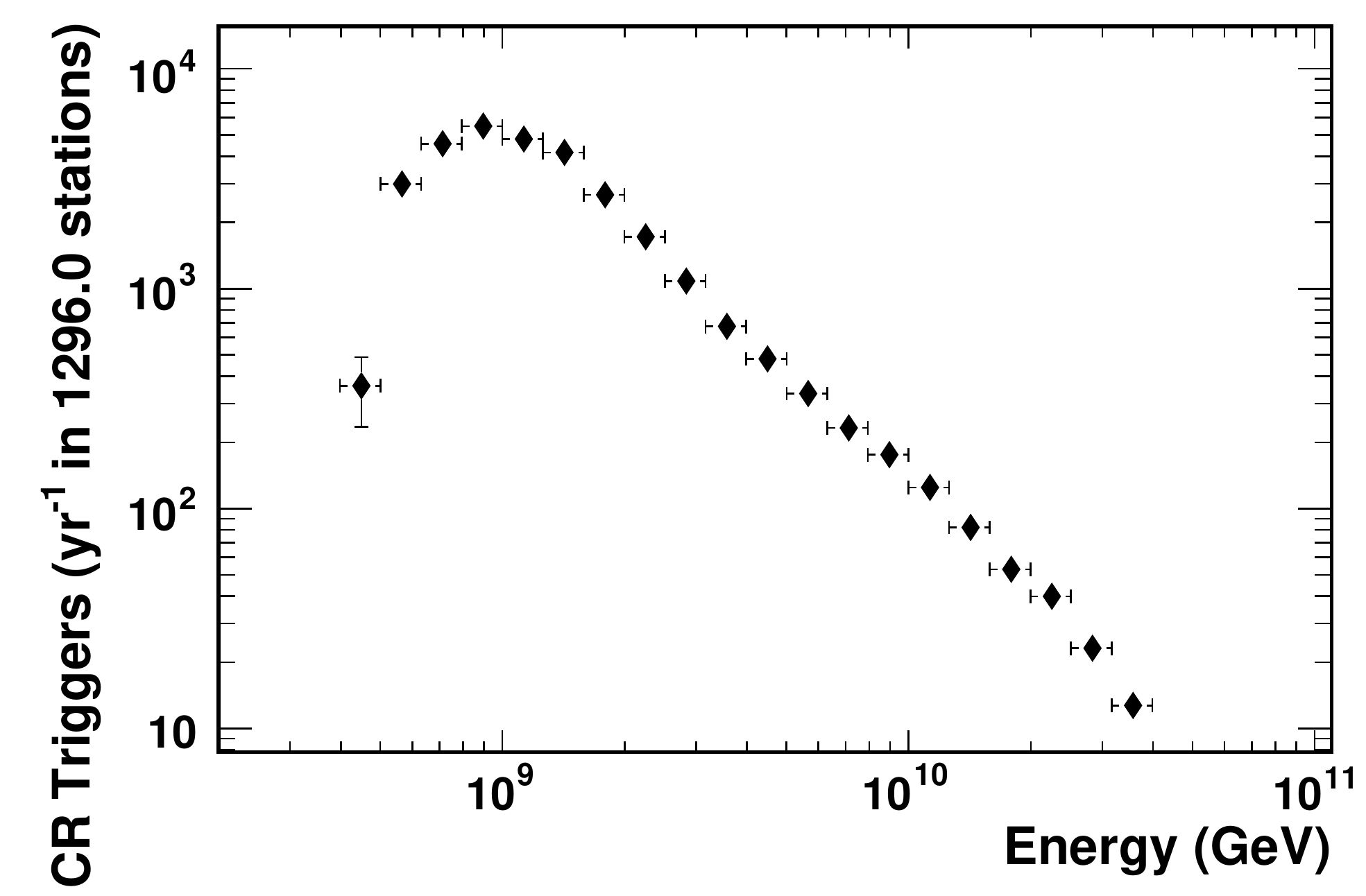}
   \end{center}
   \caption{\label{arianna:fig:CRspectrum}
     The number of triggered cosmic ray events in 1296 {\arianna}
     stations per calendar year, with a livetime of 58\% per year,
     versus the cosmic ray energy, in tenth of a decade energy bins.}
\end{figure}

The rate of cosmic ray triggers in the full {\arianna} telescope is
shown in \fig{arianna:fig:CRspectrum} as a function of cosmic ray
energy. The rates are shown as the expected number of triggers per
calendar year in each (tenth of a decade) energy bin. A detailed
measurement of the backward lobe response of the LPDA has not yet been
performed, so a conservative estimate of the cosmic ray rates has been
obtained by overestimating the gain of the antennas for signals
arriving from above the detector.

To study the reduction of this potential background, a ten LPDA
station has been simulated for the cosmic ray studies. This station
has eight downward facing antennas (see
\sect{arianna:performance}). In addition, two upward facing LPDAs have
been added to the station geometry to help discriminate between RF
pulses arriving from above or below the station. These two antennas
are oriented upward at a 45{\dg} angle relative to the surface of the
ice and do not participate in the trigger. This geometry is shown in
\fig{arianna:fig:CRChanGeo}.

\begin{figure}[t!]
   \begin{center}
     \includegraphics[width=\linewidth]{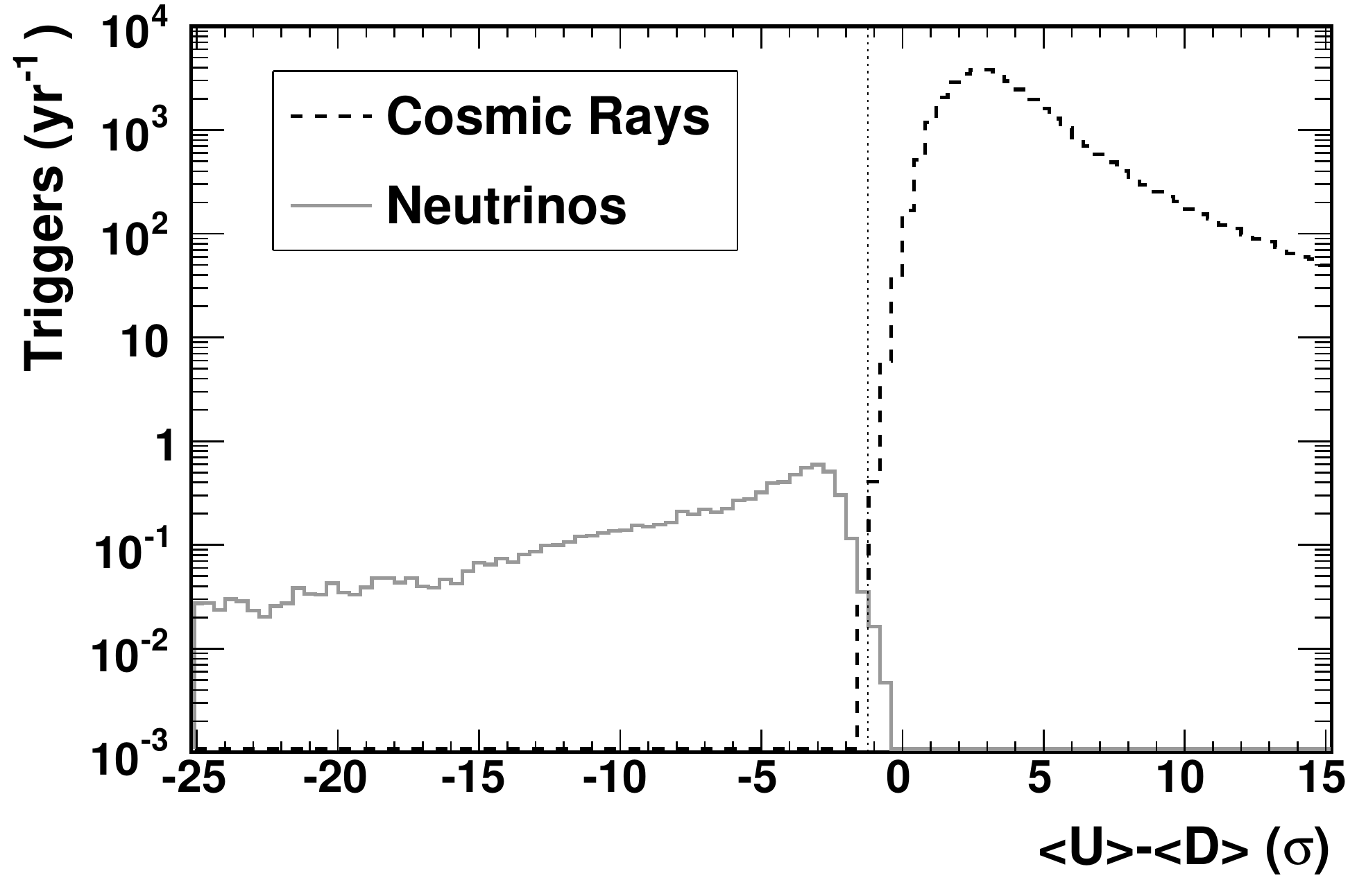}
   \end{center}
   \caption{\label{arianna:fig:cosmicRayCutEff}
   The number of triggered neutrino and cosmic ray events in 1296
   {\arianna} stations per calendar year versus the difference in the
   average amplitude of upward facing antennas and the average
   amplitude of downward facing antennas. Requiring the difference to
   be $<-1.2\sigma$ yields a background rate of $0.1$~cosmic rays in 3
   calendar years of data from the full array while preserving 99.7\%
   of the triggered neutrino events. The neutrino rate is arbitrarily
   scaled to 10~$\nu$ per year for illustration purposes.}
\end{figure}

The difference between the pulse amplitudes measured by the upward and
downward facing antennas provides an efficient mechanism for
distinguishing cosmic ray air showers from neutrino signals. On each
antenna, the amplitude is taken as the average of the two largest
crests of the time dependent waveform in order to reduce fluctuations
due to noise. \Fig{arianna:fig:cosmicRayCutEff} shows the average
upward facing antenna amplitude minus the average downward facing
antenna amplitude for both cosmic ray and neutrino events.  Keeping
only events with an amplitude difference $<\!-1.2\sigma$ leads to a
background rate of 0.1~cosmic rays in the full 1296 station {\arianna}
telescope after 3 calendar years of running with 58\% livetime per
year. This cut preserves 99.7\% of the cosmogenic neutrino events that
trigger the station.

\section{The Hexagonal Radio Array}
\label{hra}

\begin{figure}[t]
   \begin{center}
     \includegraphics[width=\linewidth]{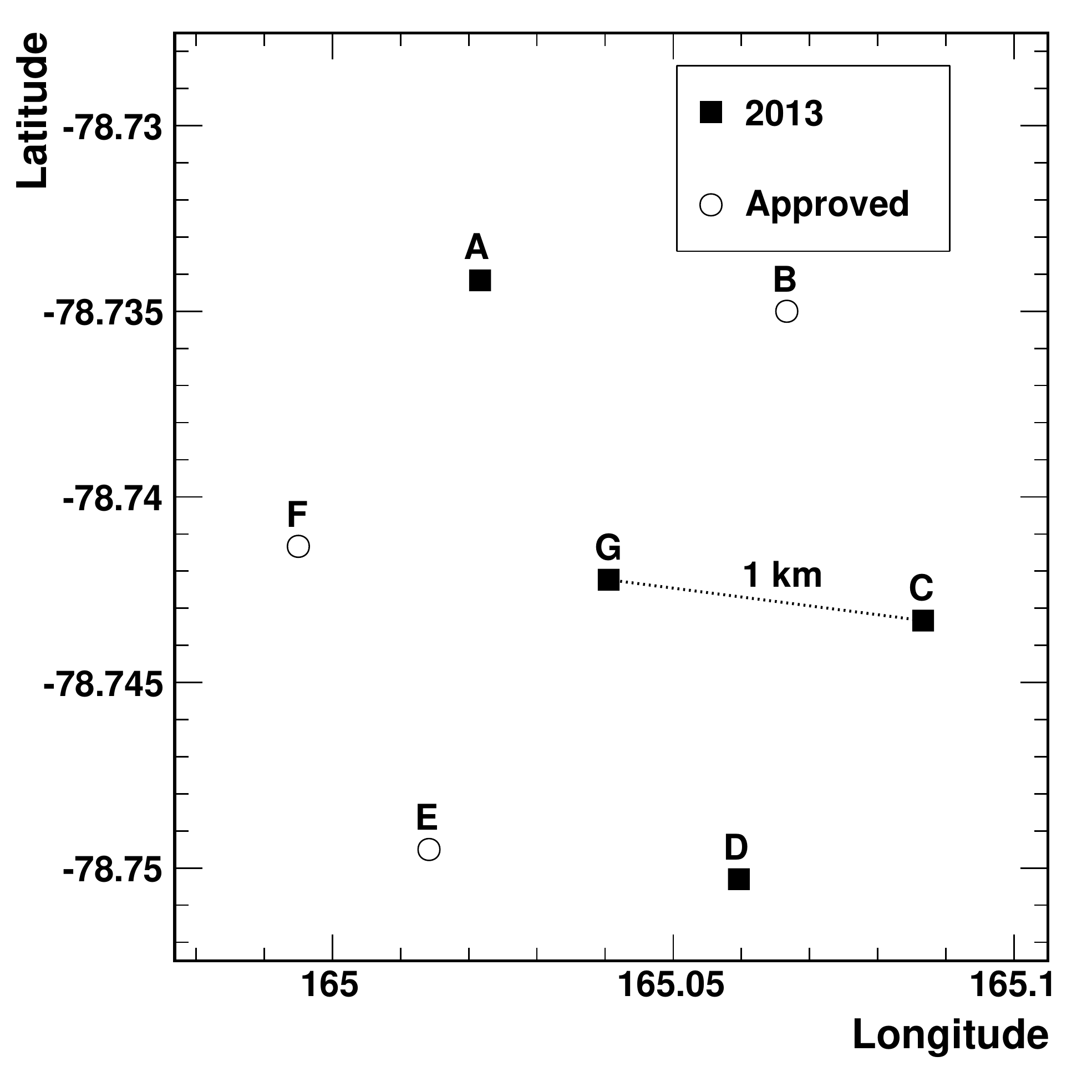}
   \end{center}
   \caption{\label{hra:fig:layout}
   The locations of the seven Hexagonal Radio Array stations. The
   HRA-3 station locations are shown by the closed squares marked
   ``A'', ``C'' and ``G'', an earlier prototype station location is
   marked ``D'', while the approved sites of the remaining four
   stations by open circles. Figure from Ref.~\cite{AriannaNIMPaper}.}
\end{figure}

The {\arianna} Hexagonal Radio Array (HRA) is being constructed on the
Ross Ice Shelf and consists of seven prototype stations arranged as
shown in \fig{hra:fig:layout}. This small array, begun in
2009~\cite{Gerhardt:2010js}, serves as a prototype for the development
and study of {\arianna} hardware, data acquisition (DAQ) and radio
data analysis. Three stations have been installed at the {\arianna}
site, Stations A, C and G, to form the HRA-3 detector. The fourth
station, Station D, is of a preliminary design installed during the
2011-2012 austral summer~\cite{sormaArianna2013}. This station did not
take radio event data during the 2013-14 season as its DAQ electronics
were removed from the station for calibration. A search has been
performed for cosmogenic neutrino signals using data taken by the
HRA-3 during the 2013-14 deployment season.

\subsection{The HRA-3 Stations}
\label{hra:hardware}

Three stations have been used to collect radio data during the 2013
deployment season. Each station consists of four downward facing LPDAs
connected to a local DAQ. The radio antennas are positioned
symmetrically around the DAQ box which lies at the center. Each
antenna is 3~{\m} from the DAQ and is oriented such that the normal
vector of the plane containing the antenna tines points toward the
DAQ. The data acquisition is able to transmit data from Antarctica in
near real-time while drawing an average of only about 7~{\watt}.

%%  The electronics draw an average of 7~{\watt} provided
%% by lithium-iron-phosphate batteries and/or solar panels.  This
%% allows each station to function independently of the others.

%% The power systems consist of two lithium-iron-phosphate batteries
%% charged by three solar panels: one 100~{\watt} panel and two
%% 30~{\watt} panels. Though the system typically requires only about
%% 5~{\watt} on average, the large capacity solar panels were deployed to
%% extend operation during the sunset and sunrise periods, and to charge
%% batteries to maintain full-time operation during long periods of
%% overcast conditions.

Radio signals measured by the LPDAs are amplified and digitized at the
data acquisition box. Signals are carried to custom amplifiers through
heavily shielded coaxial cables. The output of each amplifier is then
sampled at 1.92~{\ghz} using a custom Advanced Transient Waveform
Digitizer (ATWD) chip~\cite{sormaAtwd2013}. The chip records waveform
data in 128~samples and voltages are digitized using 12~bit analog to
digital converters.

Two complementary communication systems are used to transfer data
taken by a station to off-site locations. These systems are powered
off during data taking in order to minimize radio noise and to
conserve power. A long range wireless Ethernet link is facilitated by
an AFAR modem~\cite{AFAR} that connects to the Internet via a relay
positioned on Mt.~Discovery and a receiver at McMurdo Station. Each
station is also equipped with an Iridium Short Burst Data (SBD)
modem~\cite{SBD} that allows 320~byte binary messages to be sent via
satellite when an AFAR connection cannot be established.

Station configuration parameters, such as trigger thresholds, are
specified remotely by shift crews and transmitted to the stations
using the communication peripherals. Each station periodically
connects to computers in California in order to transmit diagnostic
data. During the connection, radio event data may be transferred and
new configuration parameters may be specified. This facilitates near
real-time data analysis and station monitoring throughout the data
taking season.

During the 2013 deployment season, events were recorded by an HRA-3
station when the time-dependent waveforms on two of four antenna
channels matched a pattern trigger. The coincidence is required to
occur within 64~{\ns}. The pattern requires the crossing of both
positive and negative $4\sigma$ thresholds and is efficient for
bipolar pulses of frequencies within the LPDA and amplifier
bandwidths. The bipolar pulse requirement reduces triggers on random
electronics noise while preserving those due to neutrino-induced
Askaryan pulses, as the finite bandwidth of the LPDA and amplifier
will always yield a bipolar pulse (due to ringing).

To facilitate studies of the thermal environment noise, the collection
of data at random times is facilitated by software forced
triggers. The stations typically record such an ``unbiased'' event
once every 67~seconds.

A detailed description of the HRA-3 power, communications and data
acquisition hardware and performance may be found in
Ref.~\cite{AriannaNIMPaper}.

\subsection{Operation of the HRA-3}
\label{hra:data}

The HRA-3 stations installed in the 2013 deployment season took data
until the lack of solar power caused the batteries to deplete in
April, 2014. The communication systems functioned as expected during
the entire data taking season. Connections over the wireless internet
link began to fail in early April due to the loss of reliable power at
the relay on Mt.~Discovery. Satellite connections over the Iridium
network continued to function until the loss of station power in
mid-April.

\begin{figure}[t]
   \begin{center}
     \includegraphics[width=\linewidth]{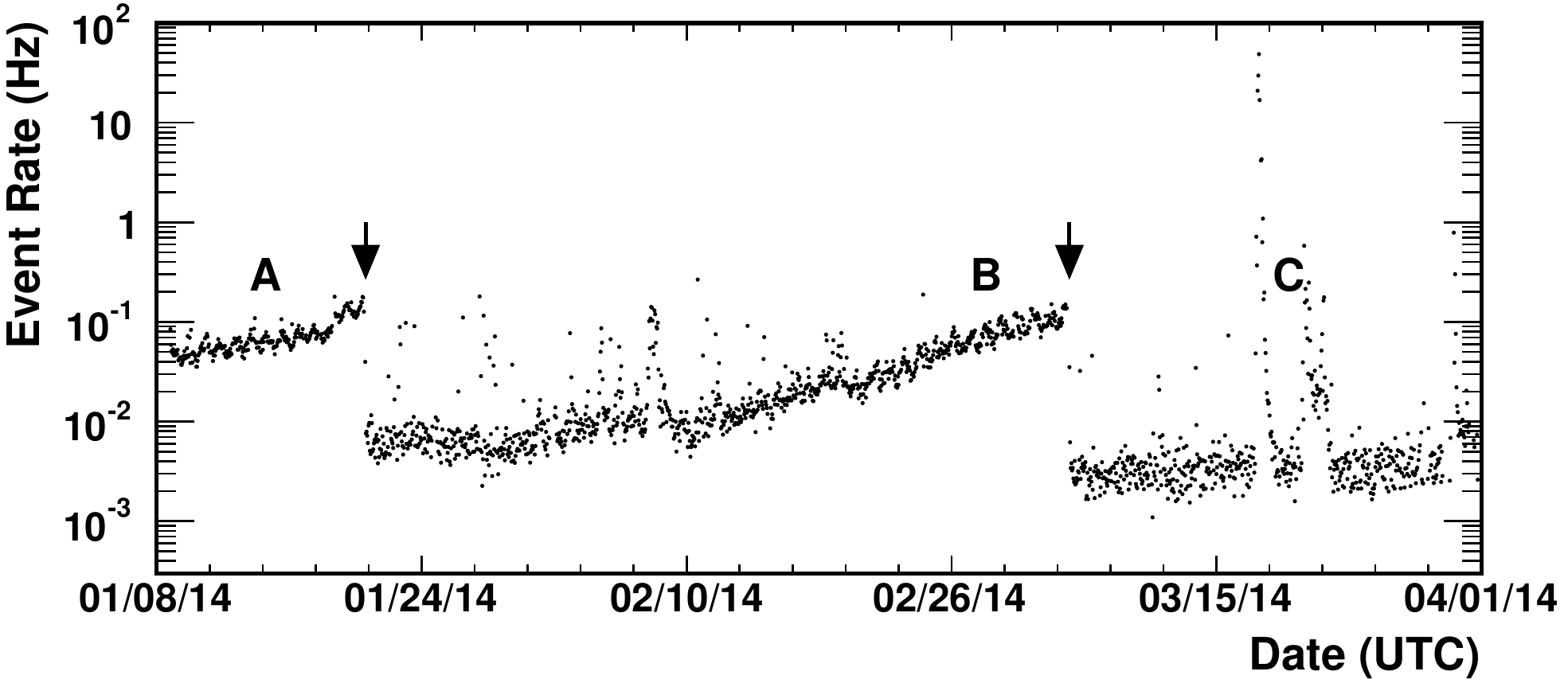}
   \end{center}
   \caption{\label{hra:fig:rateTicker}
   The rate of thermal triggers over time for Station~A. The letters
   indicate periods in which different features are seen in the
   rates. During periods A and B, both the small diurnal variations in
   rate as well as the overall gradual increase in rate is due to
   variations in the temperature of the amplifiers. The arrows
   indicate the two adjustments made to the trigger thresholds. During
   period C, a large storm passed through the {\arianna} site,
   resulting in large fluctuations of the trigger rate. Figure from
   Ref.~\cite{AriannaNIMPaper}.}
\end{figure}

The bipolar trigger requirement on two of four antenna channels,
described in \Sect{hra:hardware}, allowed the stations to run at low
trigger rates while also keeping thresholds low. As shown in
\fig{hra:fig:rateTicker}, the trigger rate of Station~A was typically
below 0.1~{\hz} and an average rate of $10^{-3}$~{\hz} has been
achieved with $4\sigma$ thresholds.

Several features in the triggering rates can be found in
\fig{hra:fig:rateTicker} during the periods marked by the letters.
Small diurnal trigger rate fluctuations seen in periods A and B are
attributed to daily temperature variations of the amplifiers. The
gradual increasing of the trigger rate during period B is caused by an
overall cooling of the amplifier electronics. This cooling raised the
gain of some amplifiers more than others, leading to different
effective thresholds on different channels and allowing small diurnal
temperature fluctuations to affect trigger rates during period B. Once
the station is fully buried in snow and thresholds are balanced, the
trigger rates are found to be stable. Such periods are observed after
each threshold tuning, marked by the arrows. Other HRA-3 stations
exhibited a similar dependence of trigger rates on temperature.

%%  Gradual increasing of the trigger
%% rate, noticeable during period B, is caused by cooling of the
%% amplifier electronics. The gain of the amplifiers increases with
%% cooler temperatures, leading to a slight increase of thermal
%% noise. Once buried, the temperature of the electronics stabilizes and
%% the baseline rate is constant. The small diurnal trigger rate
%% fluctuations seen in periods A and B are also due to temperature
%% variations of the amplifiers. The other stations experienced a similar
%% dependence of trigger rate on temperature. As the stations become
%% progressively more buried, they are insulated from temperature
%% variations by the snow cover and corresponding rate fluctuations are
%% diminished.

The typical observed thermal trigger rates match the rates expected
for the threshold values. The trigger thresholds were only adjusted
twice during the 2013-14 data taking season, denoted by the arrows in
\fig{hra:fig:rateTicker}. The threshold adjustments primarily served
to re-balance the single-channel trigger rates. This allows each
channel to participate equally in the trigger requirement that at
least two out of four antenna channels have significant bipolar
signals.

An increase in the trigger rate during storms at the {\arianna} site
has been observed. One such period, marked as period C, is visible in
\fig{hra:fig:rateTicker}. The precise cause of
these triggers continues to be investigated. During periods of high
wind speeds, above roughly 20~{\knots}, a correlated rise in rates
among all stations is seen. However, not all high wind periods result
in elevated trigger rates. The events recorded during these periods
are readily distinguishable from expected neutrino signals, as
discussed in \sect{ana}.

\section{Search for Neutrino Signals in HRA-3 Data}
\label{ana}

%% Joulien's Outline
%% -----------------

%% Procedure Overview\newline

%% Given in ADC samples from the receiver which are converted to voltages by using calibration values for this conversion are the antenna gain and the amplification of the electronics .... converted to actual field strength \newline

%% Data Set\newline

%% Station 3 - flagship ... Jan to April\newline

%% Station 10 - Jan to Feb\newline

%% Station 11 - Jan to April excluding Ch 3\newline

%% Table with breakdown of total events from all stations divided by type prior to analysis cuts.\newline

%% Runtime|LiveTime|Majority Logic Triggers|Minbias|Event Quality Cuts|Heartbeats\newline

%% Event Selection\newline

%% Describe Reject early events\newline

%% Waveform Shape Characteristics and Frequency Spectrum Data Analysis\newline

%% Describe NHM variable and cut.\newline
%% Describe average correlation coefficient variable and cut.\newline

%% Noise\newline

%% Uncertainties\newline

\subsection{The Data Set}
\label{ana:data}

Data taken by the HRA-3 between January 3 and April 9, 2014 has been
analyzed to search for neutrino-induced Askaryan signals. The former
date corresponds to the departure of the deployment crew from the Ross
Ice Shelf. The latter date is chosen to include all data successfully
transferred off of Antarctica. The entire data set taken by the HRA-3
and transferred off site is included in the analysis, resulting in a
combined livetime for the three stations of 170 days.

The bulk of the livetime deficit is not attributable to operational
deadtime. Towards the end of March, the wireless Internet connection
to the stations became less reliable, and a significant portion of the
data set taken during this period remained on the stations. With the
resumption of operations in the austral summer, the remaining data
will be transferred off site and added to the data set for
analysis. In addition, the continuous taking of radio data by
Station~G was halted during a period of severe weather at the site
just prior to midnight, February~12, 2014 (UTC). An unnecessary diode
train (not present on other stations) between the solar panel and
battery of the station failed, breaking the electrical connection
between the two. To conserve battery power, a short period each day
was used to collect radio data. The station was otherwise kept in a
low power configuration during which the battery power supply was
monitored. The station continued to operate for 3 months without
external charging. Removal of the diode train during the following
summer season allowed the station's batteries to charge.

The deadtime of each station during its normal data-taking operation
was typically 6\% or less. During such periods, each station takes
radio data continuously, pausing only for periodic off site
communications and for brief calibration data collection, as described
in \sect{hra:hardware}. The off site communications typically last for
1~minute and occur every half hour, accounting for a 4\%-5\% deadtime,
depending on the connection stability. A deadtime of 1\% results from
the collection calibration data, performed for for 10 minutes every 12
hours. These two interruptions of radio data taking constitute the
entirety of the station deadtime during normal operation. Deadtime due
to triggering and data acquisition is negligible as trigger rates are
far below the maximum acquisition rate. The intentional disabling of
radio data collection, such as the low-power operation of Station G
after its batteries were no longer able to be charged, has not been
included in the calculation of the fractional deadtime observed during
data taking.

\subsection{Neutrino Candidate Selection}
\label{ana:selection}

The search for neutrino signals in the data set has been performed
using a simple analysis for which the reconstruction of the radio
signal direction is unnecessary~\cite{JoulienPhd}. Neutrino candidate
events are required to meet three criteria. First, the event should
pass a filter designed to reject purely random thermal
triggers. Second, the event should not show evidence of electronics
noise characterized by sinusoidal waveforms. Third, to
separate neutrino candidates from non-thermal background events such
as those associated with strong winds, the waveform observed by at
least one LPDA is required to correspond reasonably well to the
waveform expected for a neutrino.

The impact of the cuts on neutrino signals has been estimated using
simulations. Neutrino events are generated according to a cosmogenic
flux energy distribution and radio signals are propagated to the
detector station in the frequency domain, as described in
\sect{arianna:sims:freq}. These simulated neutrino events are required
to pass the same trigger requirement used in the HRA-3 data: at least
2 of 4 antennas must have bipolar signals beyond 4$\sigma$. The
time-dependent waveform measured by an antenna is then determined by
first choosing the appropriate neutrino waveform template (see
\sect{arianna:sims:time}) based on the relative orientation of the
LPDA and the incoming radio signal direction. Each template is then
scaled such that its amplitude corresponds to that calculated by the
frequency-domain simulation.  Finally, finite bandwidth noise is added
to each waveform.

\begin{figure}[t]
   \begin{center}
     \includegraphics[width=\linewidth]{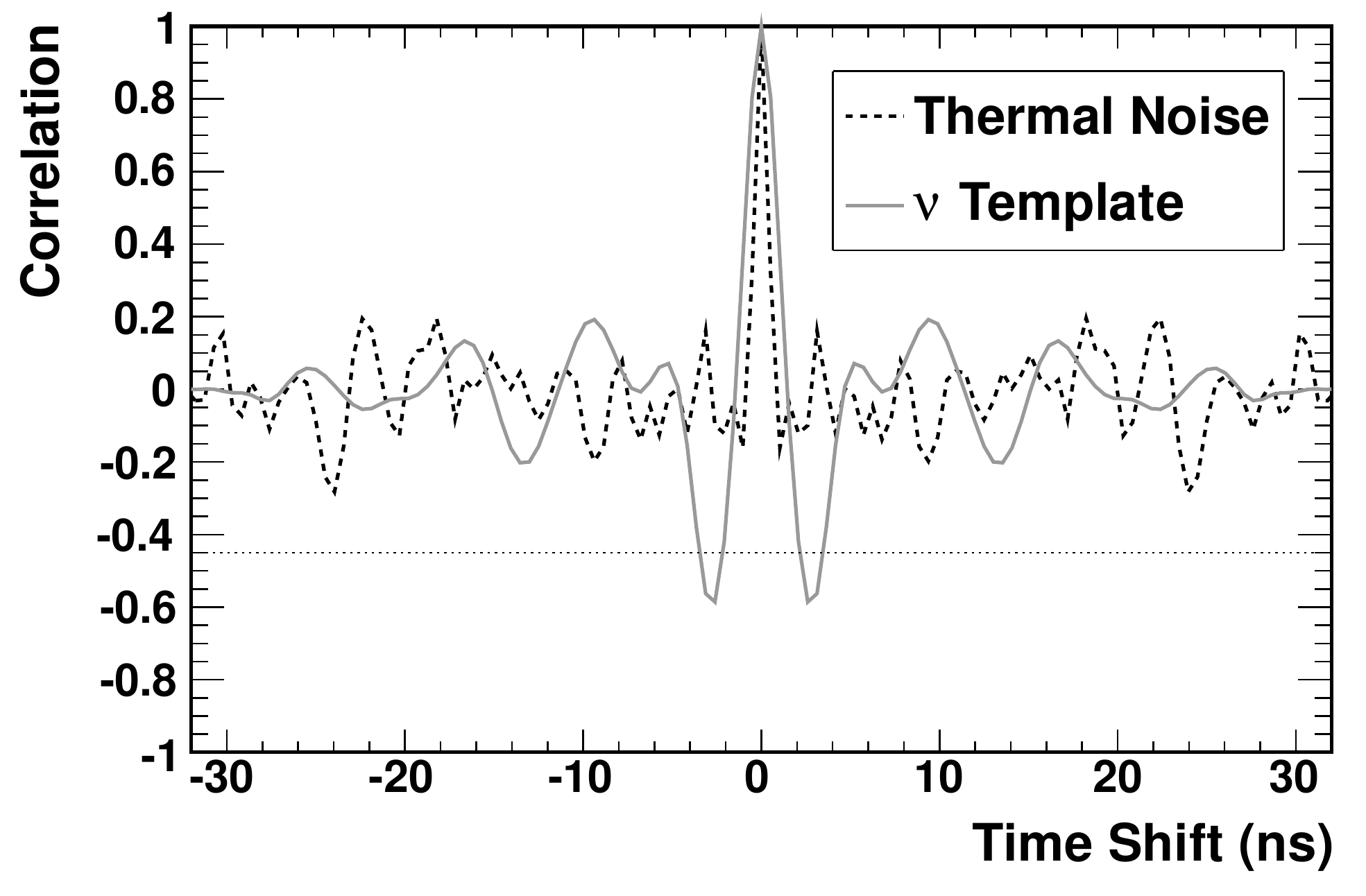}
   \end{center}
   \caption{\label{ana:fig:autoCorrComp}
     A comparison of the autocorrelation function for a purely thermal
     noise waveform (from a software forced trigger) and a neutrino
     template waveform. The dotted horizontal line shows the cut value
     ($\alpha<-0.45$) used in the analysis.}
\end{figure}

Purely random triggers are identified in the data by noting that
continuous white noise has an autocorrelation function with a perfect
correlation at zero time offset, and no correlation at other time
offsets. This property is used to distinguish purely random thermal
triggers from non-thermal events. \Fig{ana:fig:autoCorrComp} contrasts
the autocorrelation of a waveform from a software forced trigger
(purely thermal noise) with that of a waveform expected for a typical
neutrino signal.

Non-thermal events are taken to be those for which the minimum value
of the autocorrelation function, $\alpha$, falls below -0.45 on any
antenna. All other events are attributed to pure thermal noise. This
requirement correctly identifies as purely thermal 99\% of software
forced triggers in the HRA-3 data set. Of the regularly triggered
events (radio signals, thermal noise, etc.), 69\% are identified as
purely thermal. This filter is planned to be implemented locally on
the stations as a real-time ``level zero trigger.'' Such a level zero
trigger will reduce event rates sufficiently to allow the near
real-time transfer of radio event data from the full 1296 station
{\arianna} detector using only low bandwidth Iridium SBD
communications.

As expected, the autocorrelation cut rejects very few of the neutrino
signal events, with 99.5\% of the neutrino events having
$\alpha<-0.45$.

Some non-thermal events are present in the data that resemble detector
electronic noise rather than external radio noise. Such events are not
associated with external conditions like high winds. Instead, they are
characterized by strong sinusoidal waveforms and timing between
antennas that is inconsistent with a physical external signal. While
the latter property facilitates a powerful rejection based on signal
direction reconstruction, the sinusoidal structure of the waveform is
already sufficient to identify the events. These events are identified
by a strong, narrow peak in the frequency spectrum of the waveform
recorded by at least one antenna.

\begin{figure}[t]
   \begin{center}
     \includegraphics[width=\linewidth]{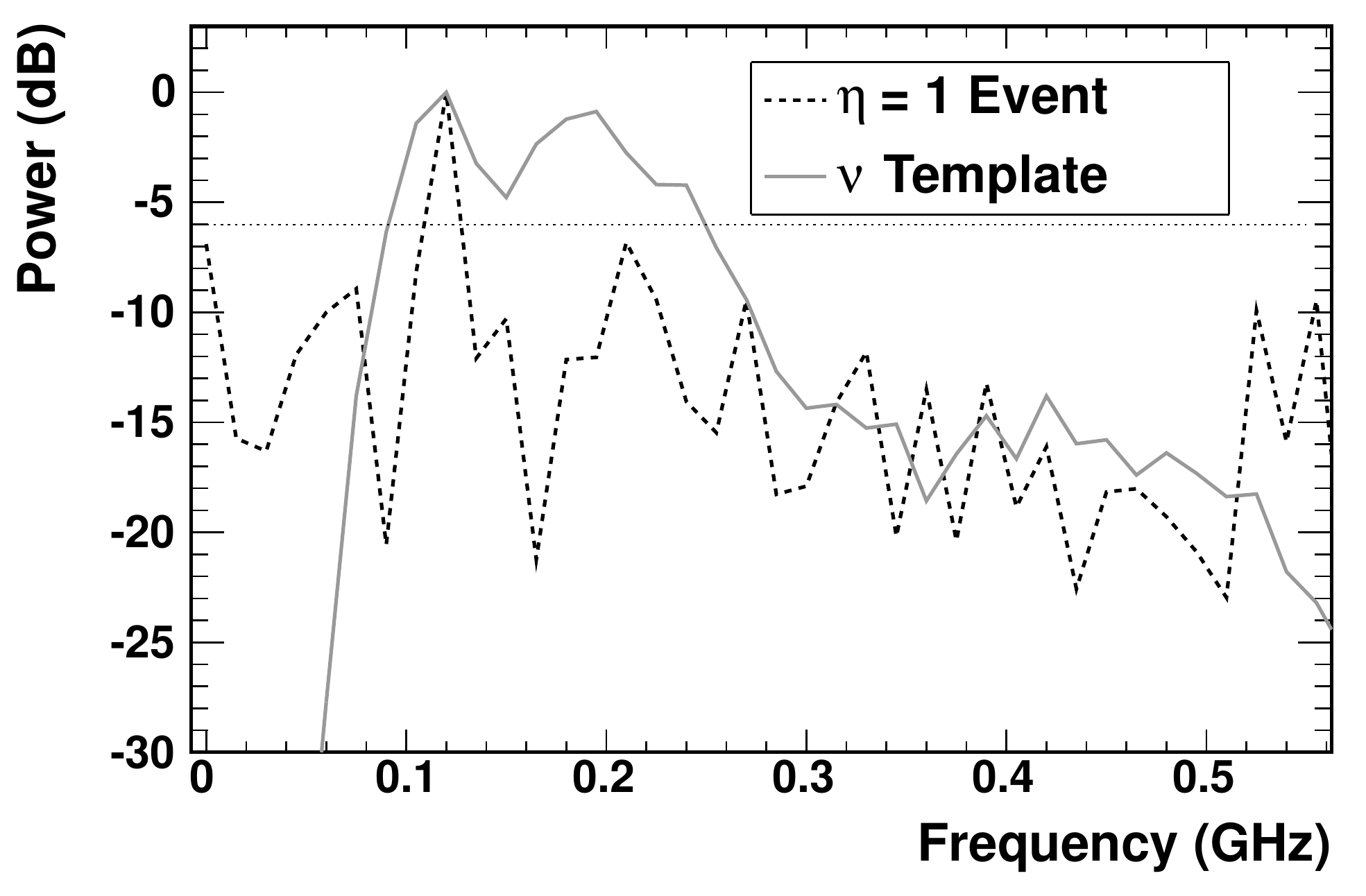}
   \end{center}
   \caption{\label{ana:fig:nhmComp}
     A comparison of the frequency spectrum of a non-thermal event
     that has a very low number of high magnitude frequency bins, in
     this case a single sharp peak at 120~{\mhz}, and a neutrino
     template waveform with no noise. Each spectrum is normalized to
     have its peak power at 0~{\db}. The dotted horizontal line shows
     the cut value used in the analysis (-6~{\db}).}
\end{figure}

\Fig{ana:fig:nhmComp} shows the frequency spectrum, measured in
10.6~{\mhz} bins, of such a sinusoidal-like waveform compared to that
expected for a typical neutrino signal. The frequency spectrum of a
neutrino candidate waveform is required to have more than 3 frequency
bins at or above half of the magnitude of the frequency bin containing
the maximum magnitude. This variable is referred to as $\eta$. It is
equivalent to the number of frequency bins containing more than one
quarter of the maximum power.

\begin{figure}[t]
   \begin{center}
     \includegraphics[width=\linewidth]{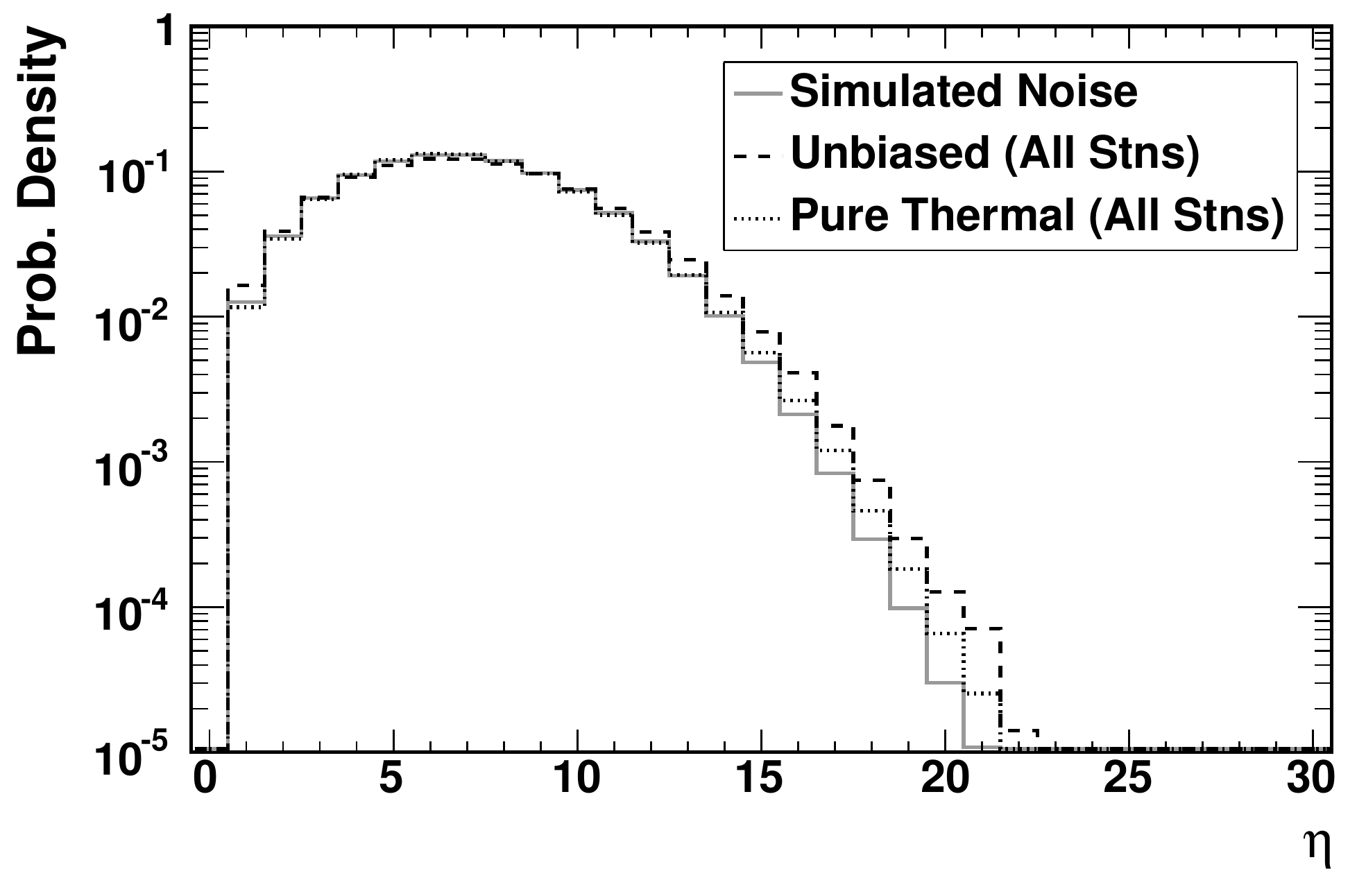}
   \end{center}
   \caption{\label{ana:fig:noiseNHM}
   Comparison of the $\eta$ distribution of pure thermal data,
   software forced (unbiased) triggered data and simulated finite
   bandwidth noise data.}
\end{figure}

The $\eta$ distribution of the thermal noise data has been compared to
the distributions in both software forced triggers as well as
simulated finite bandwidth noise, as shown in
\fig{ana:fig:noiseNHM}. This comparison shows that the cut variable
behaves as expected.

\begin{figure}[t]
   \begin{center}
     \includegraphics[width=\linewidth]{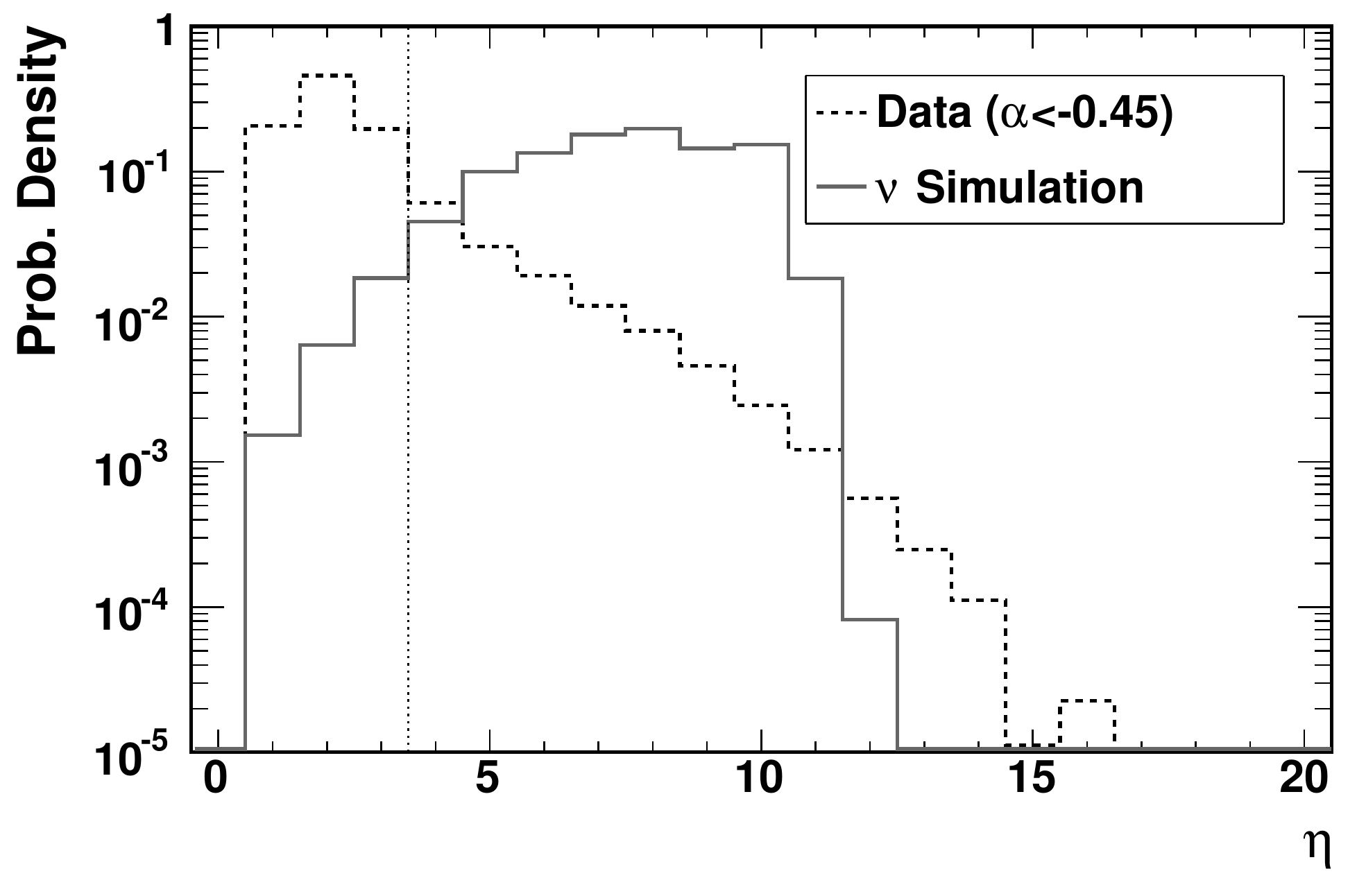}
   \end{center}
   \caption{\label{ana:fig:NHMsigbkg}
     The $\eta$ distribution of events in the data (of all stations)
     and the neutrino signal simulations. The cut, $\eta>3$, is shown
     by the dotted vertical line. Only events with $\alpha<-0.45$ are
     shown for each population.}
\end{figure}

Events with a strong frequency peak, so that $\eta\leq3$, are rejected
as being due to detector noise. This cut removes 85\% of the remaining
non-thermal triggered events across all three stations. The $\eta>3$
requirement also preserves the vast majority of neutrino signals, with
97\% of the neutrino events passing the cut. The distributions of
$\eta$ in the neutrino signal simulations and in the data for events
with $\alpha<-0.45$ are shown in \fig{ana:fig:NHMsigbkg}.

The third and final neutrino candidate selection criteria requires
that the waveform recorded in at least one antenna resemble that
expected for a neutrino signal. This is done by calculating the
maximum correlation value (for any time shift) between the waveform
reported by an antenna with a reference neutrino signal template (see
\sect{arianna:sims:time}). The relative amplitude between the waveform
recorded in the data and the neutrino template does not affect the
Pearson correlation value~\cite{correlationcoefficientsample}.

As the signal direction is not reconstructed, the relative geometry of
the signal direction and antenna orientation is not used to determine
the proper neutrino waveform template to be used as a
reference. Instead, the neutrino template corresponding to a signal
arriving with local E-plane and H-plane angles (see
\sect{arianna:sims:time}) of 30{\dg} is taken as a reference for every
antenna in every event. This reference was chosen as it represents the
average relative geometry observed in the simulations.

The calculation of the correlation between a waveform and the
reference neutrino template is complicated by the unknown relative
polarity between the recorded signal and the reference template. For
example, an LPDA with its tines oriented from East to West and its
cable connector facing South may record a bipolar pulse with a
\emph{positive} initial crest for some incoming signal. On the other
hand, rotating the LPDA by 180{\dg} so that its cable connector faces
North, with its tines oriented from West to East, would result in a
bipolar pulse with a \emph{negative} initial crest being recorded for
the same signal. To account for this effect, the correlation is
calculated for all possible unique combinations of relative polarity
between each antenna and the reference template. The best correlation
found between the reference and any antenna, across all possible
polarity orientations, is referred to as $\chi$.

\begin{figure}[t]
   \begin{center}
     \includegraphics[width=\linewidth]{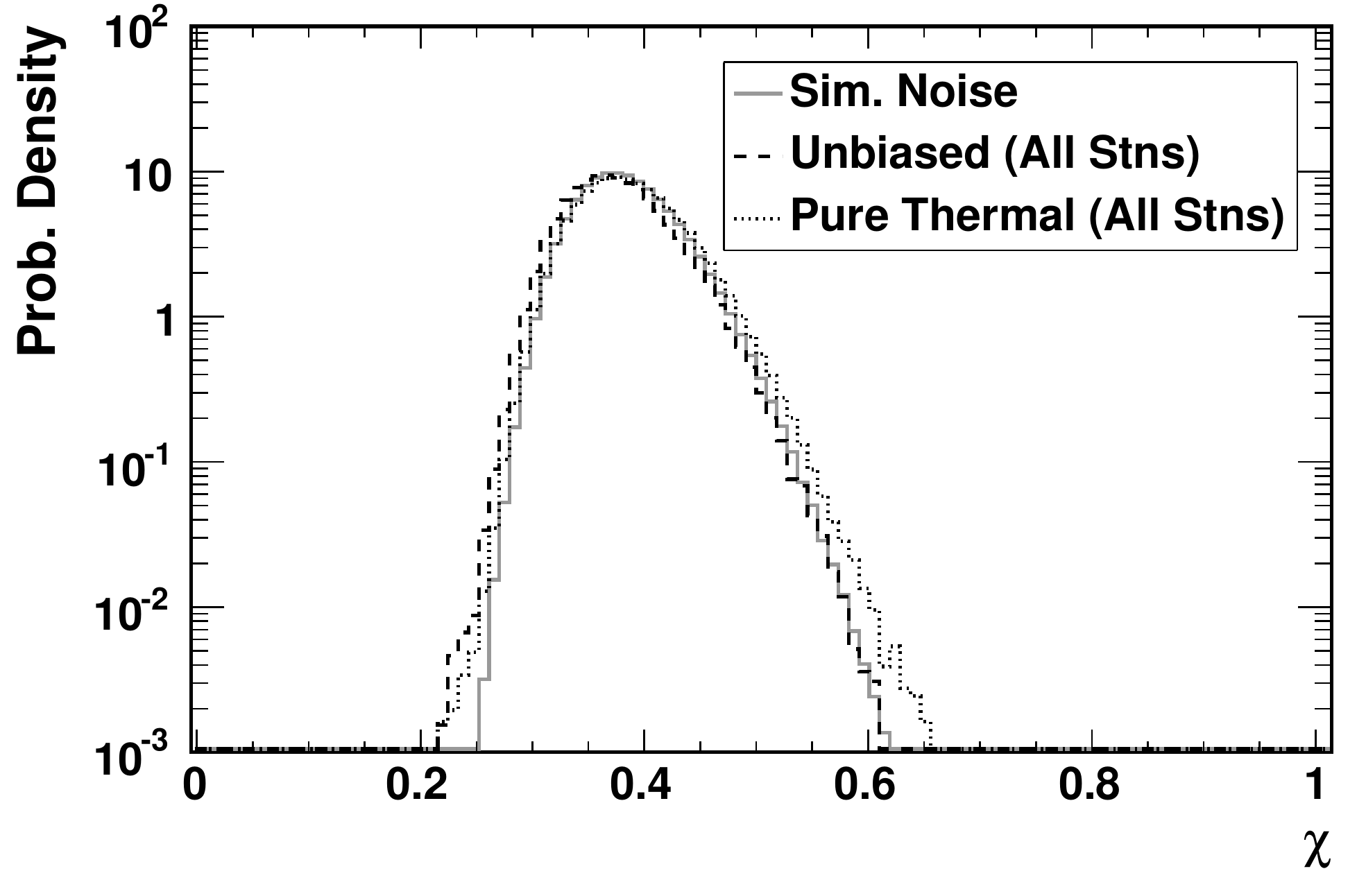}
   \end{center}
   \caption{\label{ana:fig:noiseCC}
   Comparison of the $\chi$ distribution of pure thermal data,
   software forced (unbiased) triggered data and simulated finite
   bandwidth noise data.}
\end{figure}

A comparison of the $\chi$ distribution of thermal noise data,
software forced triggers and simulated finite bandwidth noise is
shown in \fig{ana:fig:noiseCC}. This comparison shows that the $\chi$
variable behaves as expected.

\begin{figure}[t]
   \begin{center}
     \includegraphics[width=\linewidth]{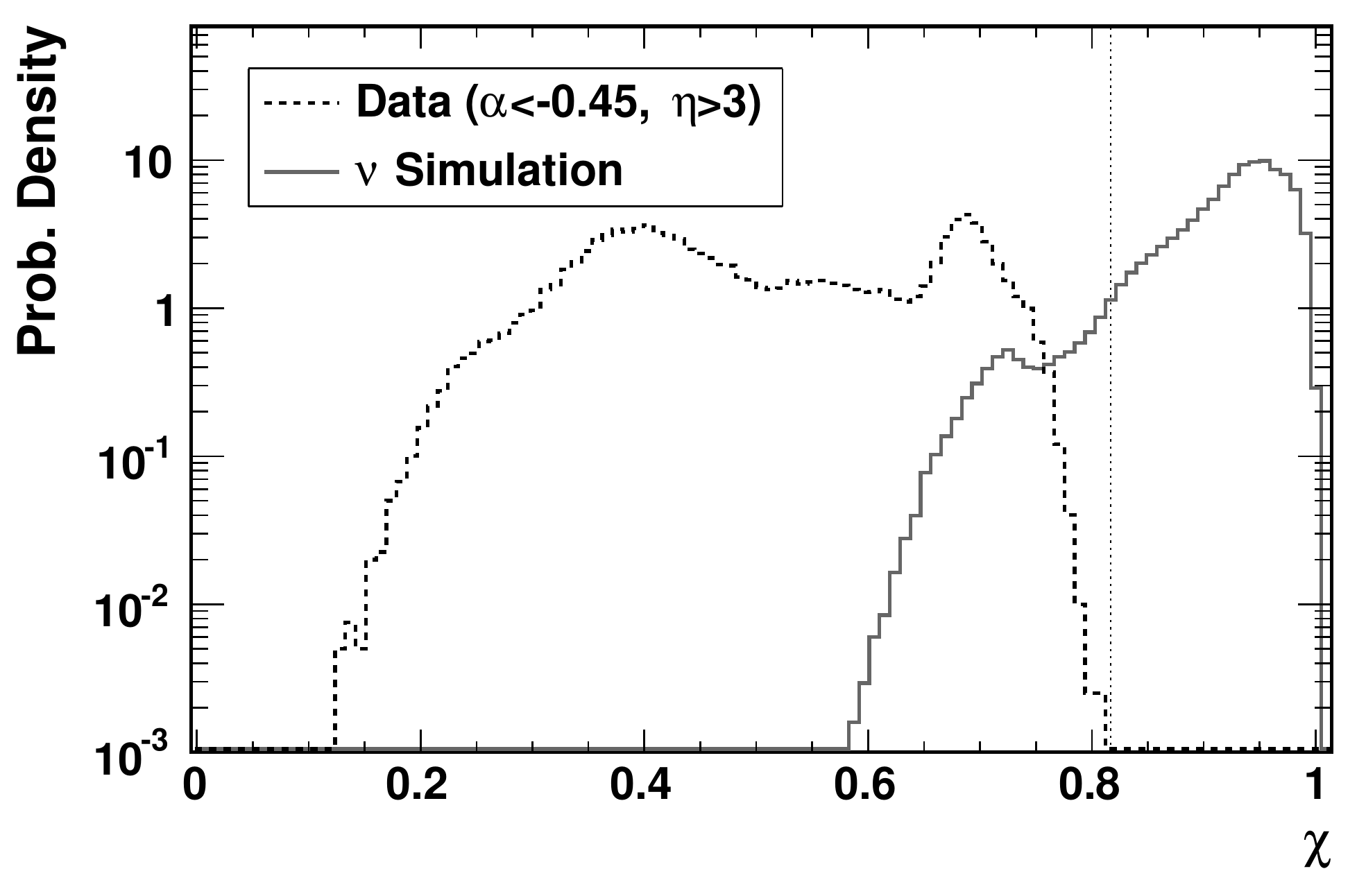}
   \end{center}
   \caption{\label{ana:fig:CCsigbkg}
     The distribution of $\chi$ in the data (of all stations) and the
     neutrino signal simulations. The cut, $\chi>0.81$, is shown by
     the dotted vertical line. Only events with both $\alpha<-0.45$
     and $\eta>3$ are shown for each population. The peak in the data
     around $\chi\approx0.68$ is due to events associated with high
     wind periods (see \fig{ana:fig:chiVsTime}).}
\end{figure}

Neutrino candidate events are required to have a $\chi>0.81$. This
seemingly large correlation value arises from the choice of cutting on
the best correlation. As shown in \fig{ana:fig:CCsigbkg}, which
compares the $\chi$ distributions of neutrino signal simulations and
events in the data having both $\alpha<-0.45$ and $\eta>3$, the cut
value is not large compared to the correlation value expected for
neutrino signals. None of the remaining events in the data set survive
the $\chi>0.81$ requirement. The requirement that $\chi>0.81$
preserves 93\% of the remaining signal events.

\begin{figure}[t]
   \begin{center}
     \includegraphics[width=\linewidth]{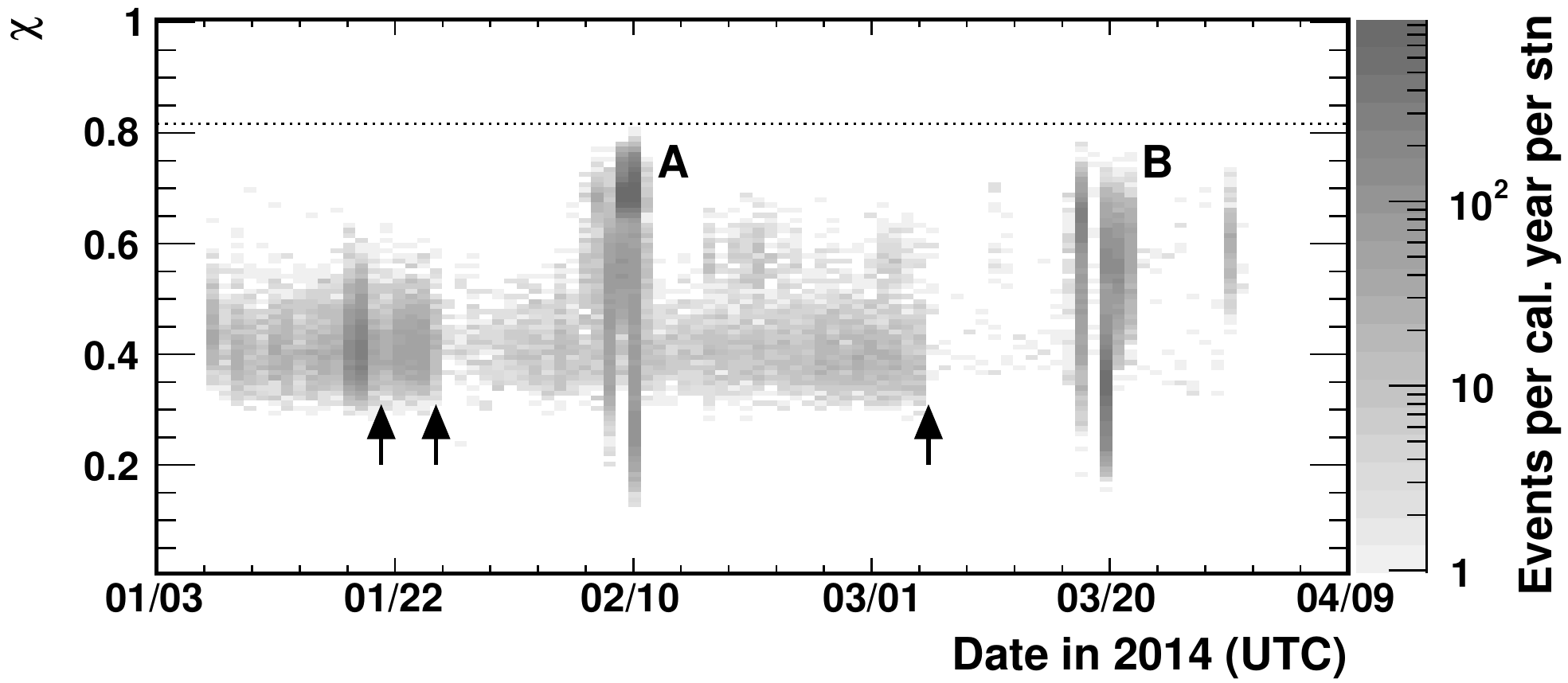}
   \end{center}
   \caption{\label{ana:fig:chiVsTime}
     The $\chi$ value of events with $\alpha<-0.45$ and $\eta>3$ on
     all three stations is shown as a function of time.  A sharp
     reduction in pure thermal events is seen after threshold tunings,
     indicated by the arrows (each station was tuned only twice, but
     not all on the same days). An excess of non-thermal background events
     is seen during the high wind periods indicated by ``A'' and
     ``B.'' The dotted line indicates the neutrino candidate cut
     ($\chi>0.81$) applied in the analysis.}
\end{figure}

\Fig{ana:fig:chiVsTime} shows the $\chi$ value as a function of time
for events with $\alpha<-0.45$ and $\eta>3$. The band of events with
$\chi\approx0.4$ is formed by purely thermal events that survive the
rather loose autocorrelation cut. Threshold tuning, indicated by the
arrows, results in a sharp reduction of these events. Although each
station had its threshold tuned only twice, different stations were
tuned at different times. An increase in non-thermal background events
is clearly seen during high wind periods, indicated by the labels
``A'' and ``B.'' While background rates increased on all stations
during these storms, Station~G recorded many more background events
than the other two stations during the storm indicated by label~A.  It
was during this storm that the batteries of Station~G stopped
receiving charge. This period accounts for both the larger number of
background events taken by Station~G as well as the peak in the data
around $\chi\approx0.68$, visible in \fig{ana:fig:CCsigbkg}.

%% The background events with the highest correlation values show further
%% properties that distinguish them from neutrino events, although these
%% properties have not been exploited. The events show a timing between
%% antennas that is inconsistent with an incoming wavefront, often with
%% pulses in three antennas showing a common phase. The pulses also show
%% strong low frequency content with a large amplitude that increases
%% over the waveform readout window. The good correlation to the neutrino
%% template arises from the low frequency content, which can resemble
%% part of the ringing pulse produced by a neutrino signal passing
%% through the finite bandwidth of the LPDA and amplifier. While these
%% events are associated with high winds, their precise sources are still
%% under investigation.

\begin{table*}[t]
   \begin{center}
      \begin{tabular}{|r|c|c|c|c|c|}
\hline
\hline
 & Station A & Station G & Station C & All Data & Cosmogenic $\nu$'s\\
\hline
Triggers & 203562 & 248772 & 512931 & 965265 & 100\%\\
\hline
$\alpha<-0.45$ & 51327 (25\%) & 102599 (41\%) & 142243 (28\%) & 296169 (31\%) & 99.5\%\\
\hline
$\eta>3$ & 3159 (2\%) & 26868 (11\%) & 13461 (3\%) & 43488 (4.5\%) & 97\%\\
\hline
$\chi>0.81$ & 0 (0\%) & 0 (0\%) & 0 (0\%) & 0 (0\%) & 90\%\\
\hline
\hline
      \end{tabular}
   \end{center}
   \caption{   \label{ana:tab:effs}
   A summary of the number of neutrino candidates remaining with the
   successive application of each cut. The fractions in parentheses
   are with respect to the totals. See text for a brief discussion of
   the excess background in the data from Station~G.}
\end{table*}

The application of all cuts preserves 90\% of the cosmogenic neutrino
triggers while removing all events recorded by the HRA-3 during the
2013-14 data taking season. A summary of the number of neutrino
candidates that remain after the successive application of each
selection criterion is presented in \tab{ana:tab:effs}.

The analysis of the 2013-14 HRA-3 data has employed a similar
procedure to that used on the HRA-3 data from the 2012-13
season~\cite{JoulienPhd} and on prototype HRA-1
data~\cite{HansonPhd}. These analyses also found high signal
efficiency values using background rejection procedures that produced
no neutrino candidate events.

\subsection{Flux Limit}
\label{ana:limit}

\begin{figure}[t]
   \begin{center}
     \includegraphics[width=\linewidth]{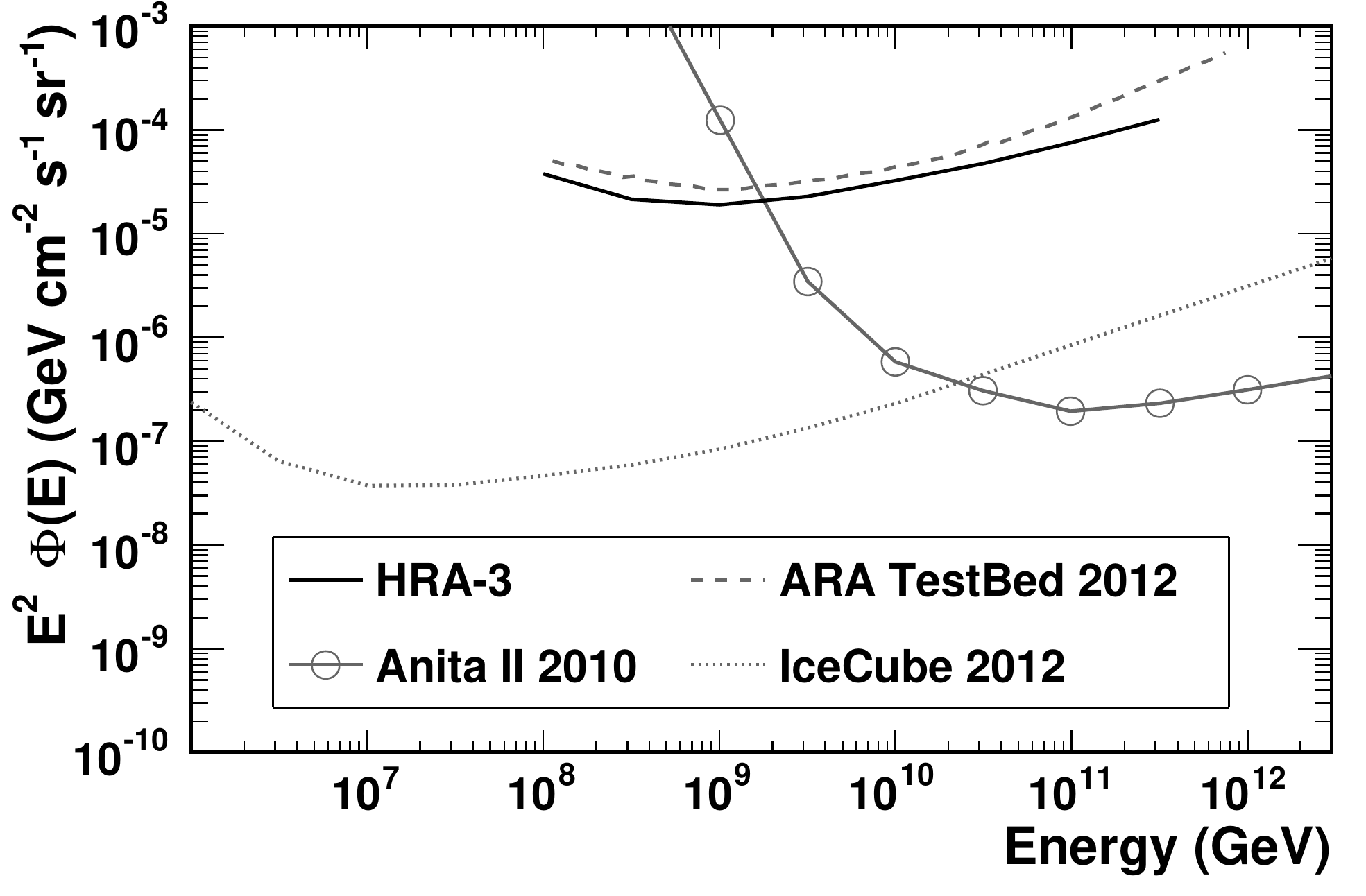}
   \end{center}
   \caption{\label{ana:fig:upperLimit} 
     The 90\% confidence level Neyman upper limit on the all flavor
     ${\nu+\bar{\nu}}$ flux, calculated in a sliding decade-wide
     energy bin, arising from the lack of neutrino candidate events in
     the HRA-3 data set collected during the 2013-14 season. The limit
     is compared to that of the ARA TestBed~\cite{Allison:2014kha} as
     well as to the current best limits set by Anita
     II~\cite{Gorham:2010kv,PhysRevD.82.022004} at high energies and
     IceCube~\cite{Aartsen:2013dsm} at lower energies.}
\end{figure}

An upper limit on the total diffuse neutrino flux can be determined
due to the absence of any observed events. With no observed events and
no background events, the Neyman formalism is used to place an upper
limit of 2.3 neutrino events at the 90\% confidence level in each
energy bin. \Fig{ana:fig:upperLimit} shows this limit translated to a
limit on the all flavor neutrino flux as a function of neutrino
energy. No neutrino flux model has been assumed in the calculation of
the limit. Instead, \eq{arianna:eq:ntrgEffvol} is used to determine
the flux that would produce 2.3 neutrinos in a sliding decade-wide
energy bin by taking $E$ to be the energy at the center of the bin,
$d\!N\leq2.3$ to be the limit on the number of neutrinos in the energy
bin and $d\!E={E\ln\!10d\!\log\!E}={E\ln\!10}$ to be the width
of the decade-wide energy bin. That is,\noindent%
\begin{equation}
   \label{ana:eq:upperLimit}
E^{2} \Phi(E) \leq \frac{2.3\ E}{\ln\!10}
 \frac{L(E)}{\varepsilon \veff\Omega\ {\tlive}}
\end{equation}
\noindent%
where $\veff\Omega$ is the single station effective volume, averaged
over neutrino flavor, shown in \fig{arianna:fig:effvol}. Also shown in
\fig{ana:fig:upperLimit} is the recent limit placed by the ARA TestBed
detector with 224 days of livetime~\cite{Allison:2014kha}, as well as
the current best limits from Anita
II~\cite{Gorham:2010kv,PhysRevD.82.022004} at high energies and
IceCube~\cite{Aartsen:2013dsm} at lower energies. Decade-wide energy
bins have been chosen to facilitate comparison with the differential
limits from ARA and IceCube.

%The lack of observation in the HRA-3 data places a limit on the all
%flavor ${\nu+\bar{\nu}}$ flux of {\fluxUpLim} at {\fluxLimEn}.

\subsection{The Full {\arianna} Detector}
\label{ana:extrap}

A search for neutrino signals in the data from the full {\arianna}
detector will make use of all information recorded by the detector in
each event. Full likelihood fits will be used to reconstruct the radio
signal direction, the incoming neutrino direction and the neutrino
energy. This will facilitate the use of further, and almost certainly
more powerful, selection criteria with which to separate neutrino
signals from radio backgrounds.

However, it is worthwhile to investigate how the simple analysis
presented in this article would perform for data taken by the full
detector. For this purpose, the full detector is taken to operate for
58\% of the year, from mid-September to early April, and to consist of
1296 stations.

The deadtime of the full detector will be reduced relative to that of
the HRA-3 stations. With the autocorrelation filter implemented as a
local level zero trigger (see \sect{ana:selection}) average event
rates of $10^{-4}$~{\hz} are easily achievable, even with $4\sigma$
thresholds. Each station can then immediately send every event off
site via the Iridium SBD satellite connection, allowing near real-time
data transfer. With such a setup, each station would transfer fewer
than 9~events per day, leading to a deadtime of 2\% due to data
transfer and communications. The calibration data will likely be
collected somewhat less often, leading to a deadtime of $<1\%$.

\begin{figure}[t]
   \begin{center}
     \includegraphics[width=\linewidth]{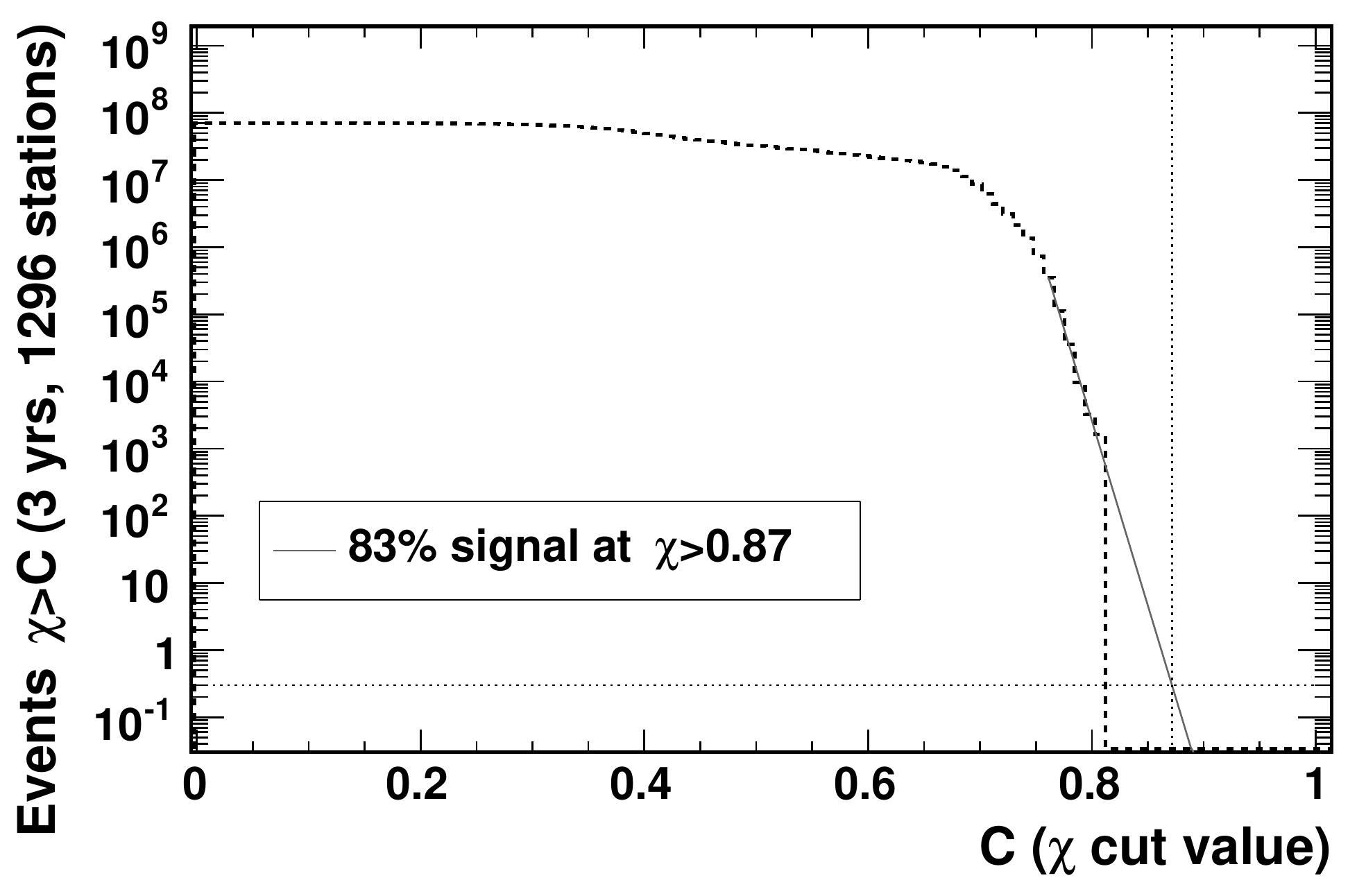}
   \end{center}
   \caption{\label{ana:fig:ccExtrap}
     The cumulative background distribution from the current HRA-3
     analysis, scaled up to a livetime equivalent to 1296 stations
     running for 58\% of 3~years. The solid grey line shows an
     exponential fit to the tail used to extrapolate the cut necessary
     to allow 0.3~background events in the full detector data
     set. This cut preserves 83\% of the cosmogenic neutrino
     triggers.}
\end{figure}

\Fig{ana:fig:ccExtrap} shows the cumulative number of background
events expected to pass a $\chi>C$ cut as a function of the cut value,
$C$. The cumulative distribution is obtained by scaling that of the
HRA-3 analysis (see \fig{ana:fig:CCsigbkg}) up to the livetime
expected for a 1296 station detector running for 58\% of 3~years. An
exponential function is fit to the tail of the cumulative distribution
in order to determine the correlation cut necessary to admit only
0.3~background events in the full detector data set. The extrapolated
correlation cut value of $\chi>0.87$ preserves 83\% of the cosmogenic
neutrino triggers. This result assumes, by necessity, that no new type
of background event will be observed that changes the shape of the
tail of the correlation cut distribution.

The signal efficiency for an analysis with the full {\arianna}
detector should further improve, given the expectation that a
reconstruction of the neutrino angle and energy will strengthen the
signal and background separation. Also note that the extrapolation of
the HRA-3 data includes background events caused by cosmic rays. The
prior rejection of cosmic rays using upward facing antennas, as
described in \sect{arianna:cosmics}, may reduce the background
further, allowing for a less strict $\chi$ cut and a correspondingly
improved signal efficiency.

%% In addition, the majority of the
%% current background triggers are associated with high wind periods, as
%% seen in \fig{ana:fig:chiVsTime}. It is reasonable to assume that with
%% time, background sources will be better understood, and more efficient
%% filters will be implemented that further improve the separation of
%% signal from background.

%
\section{Conclusions}
\label{conc}

The {\arianna} experiment proposes the use of the Askaryan effect to
search for a diffuse flux of neutrinos in the $10^{8}-10^{10}$~{\gev}
energy range. The experiment will exploit the long attenuation length
of ice at radio frequencies by populating the surface of the Ross Ice
Shelf of Antarctica with a grid of radio detectors to reach an
effective volume on the order of 100~${\km}^{3}$. The ice to water
interface at the bottom of the ice shelf acts as a mirror to radio
pulses, making {\arianna} sensitive to neutrinos arriving from above
the detector as well as from the horizon.

The response of such a detector to the electric fields generated by
the Askaryan effect has been simulated in both the frequency and time
domains. The angular resolution of the neutrino direction is found to
be about 2.9{\dg} in the local zenith angle and 2.5{\dg} in the local
azimuth angle. Determination of the angle between the {\cheren} cone
peak and the observation angle of the detector to within 1.5{\dg}
improves the angular resolution on the neutrino direction by 50\% in
zenith and 20\% in azimuth. The resolution on the energy of the
neutrino is found to be about a factor of 5 with no knowledge of the
{\cheren} observation angle, and a factor of 2.2 if the angle is known
to about 1.4{\dg} on average.

To facilitate the construction of the full {\arianna} detector, a
prototype of seven radio detector stations, the HRA, is currently
under construction at the Ross Ice Shelf. During the 2013-14 austral
summer, three such stations collected radio data. This data set has
been analyzed in a search for high energy neutrino signals. No
neutrino-like signals were found in the data. The majority of data
collected consists of purely random thermal triggers. Two other
principal sources of backgrounds were observed, one showing
oscillatory waveforms indicative of detector electronics effects and
another associated with periods of high winds at the site.

The rejection of these backgrounds by a simple analysis that does not
rely on reconstructing the radio signal direction is found to preserve
90\% of cosmogenic neutrino triggers. A model-independent differential
upper limit has been placed at the 90\% confidence level on the all
flavor ${\nu+\bar{\nu}}$ flux in a sliding decade-wide energy bin. The
limit reaches a minimum of {\fluxUpLim} in the {\fluxEnBin} energy
bin.

The analysis presented in this article has been extrapolated to the
full 1296 station {\arianna} experiment being run during the austral
summer for 3~years. Background rejection levels are found that allow
only 0.1~background events per year in the entire detector while
preserving 83\% of cosmogenic neutrino triggers. The actual
performance of the full experiment may be further improved by the
reconstruction of the neutrino direction and energy.

%% Extrapolating the analysis presented in this article to the full 1296
%% station {\arianna} experiment running during the austral summer for
%% 3~years yields background rejection levels that allow 0.1~background
%% events per year in the entire detector while preserving 83\% of
%% cosmogenic neutrino triggers. The actual performance of the full
%% experiment will be further improved by the reconstruction of the
%% neutrino direction and energy.

The {\arianna} experiment has the potential to extend the measurement
of the diffuse neutrino flux to higher neutrino energies by two orders
of magnitude in energy. The current UHE neutrino flux measurement by
the IceCube collaboration~\cite{Aartsen:2013jdh} would produce about
2.8~neutrinos in 3 calendar years of {\arianna} data, even if there is
no additional source of high energy neutrinos, as shown in
\tab{arianna:tab:models}. A measurement of the neutrino flux at such
energies would likely generate comparable levels of interest in the
growing field of ultra-high energy neutrino astronomy. Measurements of
the diffuse flux could be improved by expanding the array, while
dedicated point searches with improved angular resolution could be
performed by increasing the density of the array.

\section{Acknowledgements}
\label{ack}

The authors wish to thank the staff of Antarctic Support Contractors,
Lockheed, and the entire crew at McMurdo Station for excellent
logistical support. The authors also thank Prof. De Flaviis for the
use of the Far Field Anechoic Chamber at U.C. Irvine and would like to
acknowledge and thank the CReSIS project and the Anechoic Chamber
facility management for the use of the world class anechoic chamber at
the University of Kansas.

This work was supported by generous funding from the Office of Polar
Programs and the Physics Division of the US National Science
Foundation, including via grant awards ANT-08339133, NSF-0970175, and
NSF-1126672.
\bibliographystyle{elsarticle-num}
\biboptions{sort&compress}
\bibliography{HraScience}

\begin{thebibliography}{10}
\expandafter\ifx\csname url\endcsname\relax
  \def\url#1{\texttt{#1}}\fi
\expandafter\ifx\csname urlprefix\endcsname\relax\def\urlprefix{URL }\fi
\expandafter\ifx\csname href\endcsname\relax
  \def\href#1#2{#2} \def\path#1{#1}\fi

\bibitem{LetessierSelvon:2011dy}
A.~Letessier-Selvon, T.~Stanev, {Ultrahigh Energy Cosmic Rays}, Rev.Mod.Phys.
  83 (2011) 907--942.
\newblock \href {http://arxiv.org/abs/1103.0031} {\path{arXiv:1103.0031}},
  \href {http://dx.doi.org/10.1103/RevModPhys.83.907}
  {\path{doi:10.1103/RevModPhys.83.907}}.

\bibitem{Abreu:2012ybu}
P.~Abreu, et~al., {Constraints on the origin of cosmic rays above $10^{18}$ eV
  from large scale anisotropy searches in data of the Pierre Auger
  Observatory}, Astrophys.J. 762 (2012) L13.
\newblock \href {http://arxiv.org/abs/1212.3083} {\path{arXiv:1212.3083}},
  \href {http://dx.doi.org/10.1088/2041-8205/762/1/L13}
  {\path{doi:10.1088/2041-8205/762/1/L13}}.

\bibitem{Weiler:2000ku}
T.~J. Weiler, {Extreme energy cosmic rays: Puzzles, models, and maybe
  neutrinos}, AIP Conf.Proc. 579 (2001) 58--77.
\newblock \href {http://arxiv.org/abs/hep-ph/0103023}
  {\path{arXiv:hep-ph/0103023}}, \href {http://dx.doi.org/10.1063/1.1398161}
  {\path{doi:10.1063/1.1398161}}.

\bibitem{Sigl:2002yk}
G.~Sigl, {The Enigma of the highest energy particles of nature}, Annals Phys.
  303 (2003) 117--141.
\newblock \href {http://arxiv.org/abs/astro-ph/0210049}
  {\path{arXiv:astro-ph/0210049}}, \href
  {http://dx.doi.org/10.1016/S0003-4916(02)00021-0}
  {\path{doi:10.1016/S0003-4916(02)00021-0}}.

\bibitem{Sigl:2000sp}
G.~Sigl, {Particle and astrophysics aspects of ultrahigh-energy cosmic rays},
  Lect.Notes Phys. 556 (2000) 259--300.
\newblock \href {http://arxiv.org/abs/astro-ph/0008364}
  {\path{arXiv:astro-ph/0008364}}.

\bibitem{Greisen:1966jv}
K.~Greisen, {End to the cosmic ray spectrum?}, Phys.Rev.Lett. 16 (1966)
  748--750.
\newblock \href {http://dx.doi.org/10.1103/PhysRevLett.16.748}
  {\path{doi:10.1103/PhysRevLett.16.748}}.

\bibitem{Zatsepin:1966jv}
G.~Zatsepin, V.~Kuzmin, {Upper limit of the spectrum of cosmic rays}, JETP
  Lett. 4 (1966) 78--80.

\bibitem{Stecker:1973sy}
F.~Stecker, {Ultrahigh energy photons, electrons and neutrinos, the microwave
  background, and the universal cosmic ray hypothesis}, Astrophys.Space Sci. 20
  (1973) 47--57.
\newblock \href {http://dx.doi.org/10.1007/BF00645585}
  {\path{doi:10.1007/BF00645585}}.

\bibitem{Beresinsky:1969qj}
V.~Berezinsky, G.~Zatsepin, {Cosmic rays at ultrahigh-energies (neutrino?)},
  Phys.Lett. B28 (1969) 423--424.
\newblock \href {http://dx.doi.org/10.1016/0370-2693(69)90341-4}
  {\path{doi:10.1016/0370-2693(69)90341-4}}.

\bibitem{Berezinsky:1975zz}
V.~Berezinsky, A.~Y. Smirnov, {Cosmic neutrinos of ultra-high energies and
  detection possibility}, Astrophys.Space Sci. 32 (1975) 461--482.
\newblock \href {http://dx.doi.org/10.1007/BF00643157}
  {\path{doi:10.1007/BF00643157}}.

\bibitem{Ahrens:2003pv}
J.~Ahrens, et~al., {Search for extraterrestrial point sources of neutrinos with
  AMANDA-II}, Phys.Rev.Lett. 92 (2004) 071102.
\newblock \href {http://arxiv.org/abs/astro-ph/0309585}
  {\path{arXiv:astro-ph/0309585}}, \href
  {http://dx.doi.org/10.1103/PhysRevLett.92.071102}
  {\path{doi:10.1103/PhysRevLett.92.071102}}.

\bibitem{Aartsen:2013dsm}
M.~Aartsen, et~al., {Probing the origin of cosmic-rays with extremely high
  energy neutrinos using the IceCube Observatory}, Phys.Rev. D88 (2013) 112008.
\newblock \href {http://arxiv.org/abs/1310.5477} {\path{arXiv:1310.5477}}.

\bibitem{Halzen:2010yj}
F.~Halzen, S.~R. Klein, {IceCube: An Instrument for Neutrino Astronomy},
  Rev.Sci.Instrum. 81 (2010) 081101.
\newblock \href {http://arxiv.org/abs/1007.1247} {\path{arXiv:1007.1247}},
  \href {http://dx.doi.org/10.1063/1.3480478} {\path{doi:10.1063/1.3480478}}.

\bibitem{Gorham:2008yk}
P.~Gorham, et~al., {New Limits on the Ultra-high Energy Cosmic Neutrino Flux
  from the ANITA Experiment}, Phys.Rev.Lett. 103 (2009) 051103.
\newblock \href {http://arxiv.org/abs/0812.2715} {\path{arXiv:0812.2715}},
  \href {http://dx.doi.org/10.1103/PhysRevLett.103.051103}
  {\path{doi:10.1103/PhysRevLett.103.051103}}.

\bibitem{PhysRevD.82.022004}
P.~Gorham, et~al., Observational constraints on the ultrahigh energy cosmic
  neutrino flux from the second flight of the anita experiment, Phys. Rev. D 82
  (2010) 022004.
\newblock \href {http://dx.doi.org/10.1103/PhysRevD.82.022004}
  {\path{doi:10.1103/PhysRevD.82.022004}}.

\bibitem{Gorham:2010kv}
P.~Gorham, et~al., {Erratum: Observational Constraints on the Ultra-high Energy
  Cosmic Neutrino Flux from the Second Flight of the ANITA Experiment},
  Phys.Rev. D85 (2012) 049901.
\newblock \href {http://arxiv.org/abs/1011.5004} {\path{arXiv:1011.5004}},
  \href {http://dx.doi.org/10.1103/PhysRevD.82.022004,
  10.1103/PhysRevD.85.049901} {\path{doi:10.1103/PhysRevD.82.022004,
  10.1103/PhysRevD.85.049901}}.

\bibitem{Kravchenko:2002mm}
I.~Kravchenko, et~al., {Limits on the ultra-high energy electron neutrino flux
  from the RICE experiment}, Astropart.Phys. 20 (2003) 195--213.
\newblock \href {http://arxiv.org/abs/astro-ph/0206371}
  {\path{arXiv:astro-ph/0206371}}, \href
  {http://dx.doi.org/10.1016/S0927-6505(03)00181-6}
  {\path{doi:10.1016/S0927-6505(03)00181-6}}.

\bibitem{Kravchenko:2006qc}
I.~Kravchenko, et~al., {Rice limits on the diffuse ultrahigh energy neutrino
  flux}, Phys.Rev. D73 (2006) 082002.
\newblock \href {http://arxiv.org/abs/astro-ph/0601148}
  {\path{arXiv:astro-ph/0601148}}, \href
  {http://dx.doi.org/10.1103/PhysRevD.73.082002}
  {\path{doi:10.1103/PhysRevD.73.082002}}.

\bibitem{Abraham:2007rj}
J.~Abraham, et~al., {Upper limit on the diffuse flux of UHE tau neutrinos from
  the Pierre Auger Observatory}, Phys.Rev.Lett. 100 (2008) 211101.
\newblock \href {http://arxiv.org/abs/0712.1909} {\path{arXiv:0712.1909}},
  \href {http://dx.doi.org/10.1103/PhysRevLett.100.211101}
  {\path{doi:10.1103/PhysRevLett.100.211101}}.

\bibitem{Abraham:2009uy}
J.~Abraham, et~al., {Limit on the diffuse flux of ultra-high energy tau
  neutrinos with the surface detector of the Pierre Auger Observatory},
  Phys.Rev. D79 (2009) 102001.
\newblock \href {http://arxiv.org/abs/0903.3385} {\path{arXiv:0903.3385}},
  \href {http://dx.doi.org/10.1103/PhysRevD.79.102001}
  {\path{doi:10.1103/PhysRevD.79.102001}}.

\bibitem{Abbasi:2008hr}
{R. U. Abbasi}, et~al., {An upper limit on the electron-neutrino flux from the
  HiRes detector}, Astrophys.J. 684 (2008) 790--793.
\newblock \href {http://arxiv.org/abs/0803.0554} {\path{arXiv:0803.0554}},
  \href {http://dx.doi.org/10.1086/590335} {\path{doi:10.1086/590335}}.

\bibitem{Martens:2007ff}
K.~Martens, Hires estimates and limits for neutrino fluxes at the highest
  energies (2007).
\newblock \href {http://arxiv.org/abs/0707.4417} {\path{arXiv:0707.4417}}.

\bibitem{Aartsen:2014gkd}
M.~Aartsen, et~al., {Observation of High-Energy Astrophysical Neutrinos in
  Three Years of IceCube Data}, Phys.Rev.Lett. 113 (2014) 101101.
\newblock \href {http://arxiv.org/abs/1405.5303} {\path{arXiv:1405.5303}},
  \href {http://dx.doi.org/10.1103/PhysRevLett.113.101101}
  {\path{doi:10.1103/PhysRevLett.113.101101}}.

\bibitem{Allison:2011wk}
P.~Allison, et~al., {Design and Initial Performance of the Askaryan Radio Array
  Prototype EeV Neutrino Detector at the South Pole}, Astropart.Phys. 35 (2012)
  457--477.
\newblock \href {http://arxiv.org/abs/1105.2854} {\path{arXiv:1105.2854}},
  \href {http://dx.doi.org/10.1016/j.astropartphys.2011.11.010}
  {\path{doi:10.1016/j.astropartphys.2011.11.010}}.

\bibitem{Allison:2014kha}
P.~Allison, et~al., {First Constraints on the Ultra-High Energy Neutrino Flux
  from a Prototype Station of the Askaryan Radio Array}, Submitted to
  Astropart.Phys.\href {http://arxiv.org/abs/1404.5285}
  {\path{arXiv:1404.5285}}.

\bibitem{GNOWhitePaper}
A.~G. Vieregg, D.~Saltzberg,
  \href{http://kicp.uchicago.edu/~avieregg/gnoWhitepaper.pdf}{{Greenland
  Neutrino Observatory(GNO): Radio Detector of Ultra-high Energy Neutrinos at
  Apex Station in Greenland}} (2014).
\newline\urlprefix\url{http://kicp.uchicago.edu/~avieregg/gnoWhitepaper.pdf}

\bibitem{Gorham:2013qfa}
P.~W. Gorham, {Particle Astrophysics in NASA's Long Duration Balloon Program},
  Nucl.Phys.Proc.Suppl. 243-244 (2013) 231--238.
\newblock \href {http://dx.doi.org/10.1016/j.nuclphysbps.2013.09.012}
  {\path{doi:10.1016/j.nuclphysbps.2013.09.012}}.

\bibitem{Gorham:2011mt}
P.~Gorham, F.~Baginski, P.~Allison, K.~Liewer, C.~Miki, et~al., {The ExaVolt
  Antenna: A Large-Aperture, Balloon-embedded Antenna for Ultra-high Energy
  Particle Detection}, Astropart.Phys. 35 (2011) 242--256.
\newblock \href {http://arxiv.org/abs/1102.3883} {\path{arXiv:1102.3883}},
  \href {http://dx.doi.org/10.1016/j.astropartphys.2011.08.004}
  {\path{doi:10.1016/j.astropartphys.2011.08.004}}.

\bibitem{Askaryan:1962}
G.~A. Askaryan, JETP 14 (1962) 441.

\bibitem{Askaryan:1965}
G.~A. Askaryan, JETP 21 (1965) 658.

\bibitem{Saltzberg:2000bk}
D.~Saltzberg, P.~Gorham, D.~Walz, et~al., {Observation of the Askaryan effect:
  Coherent microwave Cherenkov emission from charge asymmetry in high-energy
  particle cascades}, Phys.Rev.Lett. 86 (2001) 2802--2805.
\newblock \href {http://arxiv.org/abs/hep-ex/0011001}
  {\path{arXiv:hep-ex/0011001}}, \href
  {http://dx.doi.org/10.1103/PhysRevLett.86.2802}
  {\path{doi:10.1103/PhysRevLett.86.2802}}.

\bibitem{Gorham:2006fy}
P.~Gorham, et~al., {Observations of the Askaryan effect in ice}, Phys.Rev.Lett.
  99 (2007) 171101.
\newblock \href {http://arxiv.org/abs/hep-ex/0611008}
  {\path{arXiv:hep-ex/0611008}}, \href
  {http://dx.doi.org/10.1103/PhysRevLett.99.171101}
  {\path{doi:10.1103/PhysRevLett.99.171101}}.

\bibitem{AlvarezMuniz:2000fw}
J.~Alvarez-Muniz, R.~Vazquez, E.~Zas, {Calculation methods for radio pulses
  from high-energy showers}, Phys.Rev. D62 (2000) 063001.
\newblock \href {http://arxiv.org/abs/astro-ph/0003315}
  {\path{arXiv:astro-ph/0003315}}, \href
  {http://dx.doi.org/10.1103/PhysRevD.62.063001}
  {\path{doi:10.1103/PhysRevD.62.063001}}.

\bibitem{AriannaIcePaper}
{S. W. Barwick}, et~al., Radio-frequency attenuation length, basal
  reflectivity, depth, and polarization measurements from moore’s bay in the
  ross ice-shelf, {Accepted for publication in J.Glaciol.}

\bibitem{0034-4885-67-10-R03}
J.~A. Dowdeswell, S.~Evans,
  \href{http://stacks.iop.org/0034-4885/67/i=10/a=R03}{Investigations of the
  form and flow of ice sheets and glaciers using radio-echo sounding}, Reports
  on Progress in Physics 67~(10) (2004) 1821.
\newline\urlprefix\url{http://stacks.iop.org/0034-4885/67/i=10/a=R03}

\bibitem{AriannaIcrc2013}
{S. W. Barwick for the ARIANNA Collaboration}, {Performance of the ARIANNA
  Prototype Array}, in: Proc. 33rd Intern. Cosmic Ray Conf., 2013.

\bibitem{Kampert:2012mx}
K.-H. Kampert, M.~Unger, {Measurements of the Cosmic Ray Composition with Air
  Shower Experiments}, Astropart.Phys. 35 (2012) 660--678.
\newblock \href {http://arxiv.org/abs/1201.0018} {\path{arXiv:1201.0018}},
  \href {http://dx.doi.org/10.1016/j.astropartphys.2012.02.004}
  {\path{doi:10.1016/j.astropartphys.2012.02.004}}.

\bibitem{Engel:2001hd}
R.~Engel, D.~Seckel, T.~Stanev, {Neutrinos from propagation of ultrahigh-energy
  protons}, Phys.Rev. D64 (2001) 093010.
\newblock \href {http://arxiv.org/abs/astro-ph/0101216}
  {\path{arXiv:astro-ph/0101216}}, \href
  {http://dx.doi.org/10.1103/PhysRevD.64.093010}
  {\path{doi:10.1103/PhysRevD.64.093010}}.

\bibitem{Seckel:2005cm}
D.~Seckel, T.~Stanev, {Neutrinos: The Key to UHE cosmic rays}, Phys.Rev.Lett.
  95 (2005) 141101.
\newblock \href {http://arxiv.org/abs/astro-ph/0502244}
  {\path{arXiv:astro-ph/0502244}}, \href
  {http://dx.doi.org/10.1103/PhysRevLett.95.141101}
  {\path{doi:10.1103/PhysRevLett.95.141101}}.

\bibitem{Connolly:2011vc}
A.~Connolly, R.~S. Thorne, D.~Waters, {Calculation of High Energy
  Neutrino-Nucleon Cross Sections and Uncertainties Using the MSTW Parton
  Distribution Functions and Implications for Future Experiments}, Phys.Rev.
  D83 (2011) 113009.
\newblock \href {http://arxiv.org/abs/1102.0691} {\path{arXiv:1102.0691}},
  \href {http://dx.doi.org/10.1103/PhysRevD.83.113009}
  {\path{doi:10.1103/PhysRevD.83.113009}}.

\bibitem{Klein:2013xoa}
S.~R. Klein, A.~Connolly, {Neutrino Absorption in the Earth, Neutrino
  Cross-Sections, and New Physics}\href {http://arxiv.org/abs/1304.4891}
  {\path{arXiv:1304.4891}}.

\bibitem{Wang:2013njo}
S.-H. Wang, P.~Chen, M.~Huang, J.~Nam, {Feasibility of Determining Diffuse
  Ultra-High Energy Cosmic Neutrino Flavor Ratio through ARA Neutrino
  Observatory}, JCAP 1311 (2013) 062.
\newblock \href {http://arxiv.org/abs/1302.1586} {\path{arXiv:1302.1586}},
  \href {http://dx.doi.org/10.1088/1475-7516/2013/11/062}
  {\path{doi:10.1088/1475-7516/2013/11/062}}.

\bibitem{Olinto:2011ng}
A.~V. Olinto, K.~Kotera, D.~Allard, {Ultrahigh Energy Cosmic Rays and
  Neutrinos}, Nucl.Phys.Proc.Suppl. 217 (2011) 231--236.
\newblock \href {http://arxiv.org/abs/1102.5133} {\path{arXiv:1102.5133}},
  \href {http://dx.doi.org/10.1016/j.nuclphysbps.2011.04.109}
  {\path{doi:10.1016/j.nuclphysbps.2011.04.109}}.

\bibitem{KamleshPhd}
K.~Dookayka, {Characterizing the Search for Ultra-High Energy Neutrinos with
  the ARIANNA Detector}, Ph.D. thesis, University of Californa, Irvine (2011).

\bibitem{Halzen:1998be}
F.~Halzen, D.~Saltzberg, {Tau-neutrino appearance with a 1000 megaparsec
  baseline}, Phys.Rev.Lett. 81 (1998) 4305--4308.
\newblock \href {http://arxiv.org/abs/hep-ph/9804354}
  {\path{arXiv:hep-ph/9804354}}, \href
  {http://dx.doi.org/10.1103/PhysRevLett.81.4305}
  {\path{doi:10.1103/PhysRevLett.81.4305}}.

\bibitem{Gandhi:1998ri}
R.~Gandhi, C.~Quigg, M.~H. Reno, I.~Sarcevic, {Neutrino interactions at
  ultrahigh-energies}, Phys.Rev. D58 (1998) 093009.
\newblock \href {http://arxiv.org/abs/hep-ph/9807264}
  {\path{arXiv:hep-ph/9807264}}, \href
  {http://dx.doi.org/10.1103/PhysRevD.58.093009}
  {\path{doi:10.1103/PhysRevD.58.093009}}.

\bibitem{Gaisser:1994yf}
T.~K. Gaisser, F.~Halzen, T.~Stanev, {Particle astrophysics with high-energy
  neutrinos}, Phys.Rept. 258 (1995) 173--236.
\newblock \href {http://arxiv.org/abs/hep-ph/9410384}
  {\path{arXiv:hep-ph/9410384}}, \href
  {http://dx.doi.org/10.1016/0370-1573(95)00003-Y}
  {\path{doi:10.1016/0370-1573(95)00003-Y}}.

\bibitem{Gandhi:1995tf}
R.~Gandhi, C.~Quigg, M.~H. Reno, I.~Sarcevic, {Ultrahigh-energy neutrino
  interactions}, Astropart.Phys. 5 (1996) 81--110.
\newblock \href {http://arxiv.org/abs/hep-ph/9512364}
  {\path{arXiv:hep-ph/9512364}}, \href
  {http://dx.doi.org/10.1016/0927-6505(96)00008-4}
  {\path{doi:10.1016/0927-6505(96)00008-4}}.

\bibitem{AlvarezMuniz:1998px}
J.~Alvarez-Muniz, E.~Zas, {The LPM effect for EeV hadronic showers in ice:
  Implications for radio detection of neutrinos}, Phys.Lett. B434 (1998)
  396--406.
\newblock \href {http://arxiv.org/abs/astro-ph/9806098}
  {\path{arXiv:astro-ph/9806098}}, \href
  {http://dx.doi.org/10.1016/S0370-2693(98)00905-8}
  {\path{doi:10.1016/S0370-2693(98)00905-8}}.

\bibitem{AlvarezMuniz:1997sh}
J.~Alvarez-Muniz, E.~Zas, {Cherenkov radio pulses from Eev neutrino
  interactions: The LPM effect}, Phys.Lett. B411 (1997) 218--224.
\newblock \href {http://arxiv.org/abs/astro-ph/9706064}
  {\path{arXiv:astro-ph/9706064}}, \href
  {http://dx.doi.org/10.1016/S0370-2693(97)01009-5}
  {\path{doi:10.1016/S0370-2693(97)01009-5}}.

\bibitem{Gerhardt:2010bj}
L.~Gerhardt, S.~R. Klein, {Electron and Photon Interactions in the Regime of
  Strong LPM Suppression}, Phys.Rev. D82 (2010) 074017.
\newblock \href {http://arxiv.org/abs/1007.0039} {\path{arXiv:1007.0039}},
  \href {http://dx.doi.org/10.1103/PhysRevD.82.074017}
  {\path{doi:10.1103/PhysRevD.82.074017}}.

\bibitem{Radioglaciology}
C.~B. V.~Bogorodsky, P.~E. Gudmandsen, Radioglaciology, Reidel Publishing Co.
  (The Netherlands), 1985.

\bibitem{schytt}
V.~Schytt, Scientific results, Norsk Polarinstitutt {(Norwegian-British-Swedish
  Antarctic Expedition)} (1958) 113--151.

\bibitem{Barrella:2010vs}
T.~Barrella, S.~W. Barwick, D.~Saltzberg, {Ross Ice Shelf in situ
  radio-frequency ice attenuation}, J.Glaciol. 57 (2011) 61--66.
\newblock \href {http://arxiv.org/abs/1011.0477} {\path{arXiv:1011.0477}},
  \href {http://dx.doi.org/10.3189/002214311795306691}
  {\path{doi:10.3189/002214311795306691}}.

\bibitem{AriannaTemplatePaper}
S.~Barwick, E.~Berg, D.~Besson, T.~Duffin, J.~Hanson, et~al., {Time Domain
  Response of the ARIANNA Detector}, Astropart.Phys. 62 (2014) 139--151.
\newblock \href {http://arxiv.org/abs/1406.0820} {\path{arXiv:1406.0820}},
  \href {http://dx.doi.org/10.1016/j.astropartphys.2014.09.002}
  {\path{doi:10.1016/j.astropartphys.2014.09.002}}.

\bibitem{AlvarezMuniz:2010ty}
J.~Alvarez-Muniz, A.~Romero-Wolf, E.~Zas, {Cherenkov radio pulses from
  electromagnetic showers in the time-domain}, Phys.Rev. D81 (2010) 123009.
\newblock \href {http://arxiv.org/abs/1002.3873} {\path{arXiv:1002.3873}},
  \href {http://dx.doi.org/10.1103/PhysRevD.81.123009}
  {\path{doi:10.1103/PhysRevD.81.123009}}.

\bibitem{AlvarezMuniz:2011ya}
J.~Alvarez-Muniz, A.~Romero-Wolf, E.~Zas, {Practical and accurate calculations
  of Askaryan radiation}, Phys.Rev. D84 (2011) 103003.
\newblock \href {http://arxiv.org/abs/1106.6283} {\path{arXiv:1106.6283}},
  \href {http://dx.doi.org/10.1103/PhysRevD.84.103003}
  {\path{doi:10.1103/PhysRevD.84.103003}}.

\bibitem{Abreu:2012zz}
P.~Abreu, et~al., {Search for point-like sources of ultra-high energy neutrinos
  at the Pierre Auger Observatory and improved limit on the diffuse flux of tau
  neutrinos}, Astrophys.J. 755 (2012) L4.
\newblock \href {http://arxiv.org/abs/1210.3143} {\path{arXiv:1210.3143}},
  \href {http://dx.doi.org/10.1088/2041-8205/755/1/L4}
  {\path{doi:10.1088/2041-8205/755/1/L4}}.

\bibitem{Abbasi:2011ji}
R.~Abbasi, et~al., {Constraints on the Extremely-high Energy Cosmic Neutrino
  Flux with the IceCube 2008-2009 Data}, Phys.Rev. D83 (2011) 092003.
\newblock \href {http://arxiv.org/abs/1103.4250} {\path{arXiv:1103.4250}},
  \href {http://dx.doi.org/10.1103/PhysRevD.84.079902,
  10.1103/PhysRevD.83.092003} {\path{doi:10.1103/PhysRevD.84.079902,
  10.1103/PhysRevD.83.092003}}.

\bibitem{Kravchenko:2011im}
I.~Kravchenko, S.~Hussain, D.~Seckel, D.~Besson, E.~Fensholt, et~al., {Updated
  Results from the RICE Experiment and Future Prospects for Ultra-High Energy
  Neutrino Detection at the South Pole}, Phys.Rev. D85 (2012) 062004.
\newblock \href {http://arxiv.org/abs/1106.1164} {\path{arXiv:1106.1164}},
  \href {http://dx.doi.org/10.1103/PhysRevD.85.062004}
  {\path{doi:10.1103/PhysRevD.85.062004}}.

\bibitem{Aartsen:2013jdh}
M.~Aartsen, et~al., {Evidence for High-Energy Extraterrestrial Neutrinos at the
  IceCube Detector}, Science 342 (2013) 1242856.
\newblock \href {http://arxiv.org/abs/1311.5238} {\path{arXiv:1311.5238}},
  \href {http://dx.doi.org/10.1126/science.1242856}
  {\path{doi:10.1126/science.1242856}}.

\bibitem{Waxman:1998yy}
E.~Waxman, J.~N. Bahcall, {High-energy neutrinos from astrophysical sources: An
  Upper bound}, Phys.Rev. D59 (1999) 023002.
\newblock \href {http://arxiv.org/abs/hep-ph/9807282}
  {\path{arXiv:hep-ph/9807282}}, \href
  {http://dx.doi.org/10.1103/PhysRevD.59.023002}
  {\path{doi:10.1103/PhysRevD.59.023002}}.

\bibitem{Yuksel:2006qb}
H.~Yuksel, M.~D. Kistler, {Enhanced cosmological GRB rates and implications for
  cosmogenic neutrinos}, Phys.Rev. D75 (2007) 083004.
\newblock \href {http://arxiv.org/abs/astro-ph/0610481}
  {\path{arXiv:astro-ph/0610481}}, \href
  {http://dx.doi.org/10.1103/PhysRevD.75.083004}
  {\path{doi:10.1103/PhysRevD.75.083004}}.

\bibitem{Kotera:2010yn}
K.~Kotera, D.~Allard, A.~V. Olinto, {Cosmogenic Neutrinos: parameter space and
  detectability from PeV to ZeV}, JCAP 1010 (2010) 013.
\newblock \href {http://arxiv.org/abs/1009.1382} {\path{arXiv:1009.1382}},
  \href {http://dx.doi.org/10.1088/1475-7516/2010/10/013}
  {\path{doi:10.1088/1475-7516/2010/10/013}}.

\bibitem{Yoshida:1993pt}
S.~Yoshida, M.~Teshima, {Energy spectrum of ultrahigh-energy cosmic rays with
  extragalactic origin}, Prog.Theor.Phys. 89 (1993) 833--845.
\newblock \href {http://dx.doi.org/10.1143/PTP.89.833}
  {\path{doi:10.1143/PTP.89.833}}.

\bibitem{Ahlers:2010fw}
M.~Ahlers, L.~Anchordoqui, M.~Gonzalez-Garcia, F.~Halzen, S.~Sarkar, {GZK
  Neutrinos after the Fermi-LAT Diffuse Photon Flux Measurement},
  Astropart.Phys. 34 (2010) 106--115.
\newblock \href {http://arxiv.org/abs/1005.2620} {\path{arXiv:1005.2620}},
  \href {http://dx.doi.org/10.1016/j.astropartphys.2010.06.003}
  {\path{doi:10.1016/j.astropartphys.2010.06.003}}.

\bibitem{Ave:2004uj}
M.~Ave, N.~Busca, A.~V. Olinto, A.~A. Watson, T.~Yamamoto, {Cosmogenic
  neutrinos from ultra-high energy nuclei}, Astropart.Phys. 23 (2005) 19--29.
\newblock \href {http://arxiv.org/abs/astro-ph/0409316}
  {\path{arXiv:astro-ph/0409316}}, \href
  {http://dx.doi.org/10.1016/j.astropartphys.2004.11.001}
  {\path{doi:10.1016/j.astropartphys.2004.11.001}}.

\bibitem{Fang:2013vla}
K.~Fang, K.~Kotera, K.~Murase, A.~V. Olinto, {A decisive test for the young
  pulsar origin of ultrahigh energy cosmic rays with IceCube. }\href
  {http://arxiv.org/abs/1311.2044} {\path{arXiv:1311.2044}}.

\bibitem{AriannaNIMPaper}
S.~W. Barwick, et~al., {Design and Performance of the ARIANNA Hexagonal Radio
  Array Systems}, {Submitted to IEEE Trans.Nucl.Sci.}

\bibitem{ReedIcrc2013}
{C. Reed, for the ARIANNA Collaboration}, {Performance of the ARIANNA Neutrino
  Telescope Stations}, in: Proc. 33rd Intern. Cosmic Ray Conf., 2013.

\bibitem{Huege:2013vt}
T.~Huege, M.~Ludwig, C.~W. James, {Simulating radio emission from air showers
  with CoREAS}\href {http://arxiv.org/abs/1301.2132} {\path{arXiv:1301.2132}}.

\bibitem{CoReasICRC2013}
T.~Huege, {The Renaissance of Radio Detection of Cosmic Rays}, in: Proc. 33rd
  Intern. Cosmic Ray Conf., 2013, highlight Contribution.
\newblock \href {http://arxiv.org/abs/1310.6927} {\path{arXiv:1310.6927}}.

\bibitem{Huege:2013yra}
T.~Huege, C.~W. James, {Full Monte Carlo simulations of radio emission from
  extensive air showers with CoREAS. }\href {http://arxiv.org/abs/1307.7566}
  {\path{arXiv:1307.7566}}.

\bibitem{Heck:1998vt}
D.~Heck, G.~Schatz, T.~Thouw, J.~Knapp, J.~Capdevielle, {CORSIKA: A Monte Carlo
  code to simulate extensive air showers}.

\bibitem{Ostapchenko:2005nj}
S.~Ostapchenko, {Non-linear screening effects in high energy hadronic
  interactions}, Phys.Rev. D74 (2006) 014026.
\newblock \href {http://arxiv.org/abs/hep-ph/0505259}
  {\path{arXiv:hep-ph/0505259}}, \href
  {http://dx.doi.org/10.1103/PhysRevD.74.014026}
  {\path{doi:10.1103/PhysRevD.74.014026}}.

\bibitem{Abreu:2011pj}
P.~Abreu, et~al., {The Pierre Auger Observatory I: The Cosmic Ray Energy
  Spectrum and Related Measurements. }\href {http://arxiv.org/abs/1107.4809}
  {\path{arXiv:1107.4809}}.

\bibitem{Gerhardt:2010js}
{Gerhardt, Lisa and others}, {A prototype station for ARIANNA: a detector for
  cosmic neutrinos}, Nucl.Instrum.Meth. A624 (2010) 85--91.
\newblock \href {http://arxiv.org/abs/1005.5193} {\path{arXiv:1005.5193}},
  \href {http://dx.doi.org/10.1016/j.nima.2010.09.032}
  {\path{doi:10.1016/j.nima.2010.09.032}}.

\bibitem{sormaArianna2013}
{S. Kleinfelder for the ARIANNA Collaboration}, Design and performance of the
  autonomous data acquisition system for the arianna high energy neutrino
  detector, Nuclear Science, IEEE Transactions on 60~(2) (2013) 612--618.
\newblock \href {http://dx.doi.org/10.1109/TNS.2013.2252365}
  {\path{doi:10.1109/TNS.2013.2252365}}.

\bibitem{sormaAtwd2013}
S.~Kleinfelder, S.~Chiang, W.~Huang, Multi-{GHz} synchronous waveform
  acquisition with real-time pattern-matching trigger generation, Nuclear
  Science, IEEE Transactions on 60~(5) (2013) 3785--3792.
\newblock \href {http://dx.doi.org/10.1109/TNS.2013.2279660}
  {\path{doi:10.1109/TNS.2013.2279660}}.

\bibitem{AFAR}
Model AR-24027E, AFAR Communications Inc., Santa Barbara, CA (2008).

\bibitem{SBD}
Model 9602-N, NAL Research Corp., Manassas, VA (2010).

\bibitem{JoulienPhd}
J.~E. Tatar, {Performance of Sub-Array of ARIANNA Detector Stations during
  First Year of Operation}, Ph.D. thesis, University of Californa, Irvine
  (2013).

\bibitem{correlationcoefficientsample}
G.~Upton, I.~Cook, A Dictionary of Statistics, 2nd Edition, Oxford University
  Press, 2008.

\bibitem{HansonPhd}
J.~C. Hanson, {The Performance and Initial Results of the ARIANNA Prototype},
  Ph.D. thesis, University of Californa, Irvine (2013).

\end{thebibliography}

\end{document}